\newcommand{\PRformat}{} 
\ifdefined\PRformat
\documentclass[usenatbib, twocolumn, nofootinbib, superscriptaddress, longbibliography, reprint, prl]{revtex4-1}
\else
\documentclass[usenatbib, twocolumn, nofootinbib]{aastex61}
\fi

\usepackage{multirow}
\usepackage{natbib}
\usepackage{color}
\usepackage{amsmath,amsfonts,bm}
\usepackage{hyperref}
\usepackage[T1]{fontenc}
\usepackage{graphicx}	

\ifdefined\PRformat
\else
\newcolumntype{C}[1]{>{\centering\let\newline\\\arraybackslash\hspace{0pt}}m{#1}}
\fi

\ifdefined\PRformat

\newcommand{\aap}{A\&A}
\newcommand{\araa}{Annual Review of A\&A}
\newcommand{\pasp}{PASP} 

\newcommand{\apjs}{\apj S} 

\newcommand\mnras{MNRAS}

\newcommand\jcap{JCAP}
\newcommand\physrep{Physics Reports}
\fi
\newcommand{\sptpol}{SPTpol} 
\newcommand{\planck}{{\it Planck}}
\newcommand{\snr}{$S/N$}

\newcommand{\QU}{$Q/U$}

\newcommand{\howmanyclustersaboverichtwenty}{3,868}
\newcommand{\howmanyclustersaboverichten}{17,661}
\newcommand{\howmanyclustersaboverichtenfornulltest}{18,981}

\newcommand{\howmanysigmastackedforpolaboverichten}{4.8}
\newcommand{\howmanysigmastackedforpolaboverichtwenty}{4.1}

\newcommand{\desrmfullsamplemeanmassabovetennoerrors}{$0.96 \munits$} 
\newcommand{\desrmfullsamplemeanmassabovetensims}{$0.94 \pm 0.07 \munits$} 

\newcommand{\desrmfullsamplemeanmasspolsysaboverichten}{\ensuremath{(1.43 \pm 0.40) \munits}}

\newcommand{\desrmfullsamplemeanmasspolsysaboverichtennounits}{\ensuremath{1.43 \pm 0.40}}
\newcommand{\desrmfullsamplemeanmasspolsysaboverichtwentynounits}{\ensuremath{3.23 \pm 1.01}}
\newcommand{\desrmfullsamplemeanmasstempaboveten}{\mbox{$0.85 \pm 0.16$}}
\newcommand{\desrmfullsamplemeanmassaboveten}{\mbox{$0.96 \pm 0.07$}}
\newcommand{\desrmfullsamplemeanmasstempabovetwenty}{\mbox{$1.80 \pm 0.33$}}
\newcommand{\desrmfullsamplemeanmassabovetwenty}{\mbox{$2.06 \pm 0.14$}}

\newcommand{\desrmfullsamplemeanmasspolsysaboverichtenforQminusUnulltest}{\ensuremath{(-0.51 \pm 0.57) \munits}}
\newcommand{\desrmfullsamplemeanmasspolsysaboverichtenforQ}{\ensuremath{(1.30 \pm 0.57) \munits}}
\newcommand{\desrmfullsamplemeanmasspolsysaboverichtenforU}{\ensuremath{(1.56 \pm 0.54) \munits}}

\newcommand{\desrmfullsamplemeanmasspolsysaboverichtenulltest}{\ensuremath{(0.15 \pm 0.39) \munits}}

\newcommand{\datavec}{\textbf{\textrm{d}}}
\newcommand{\datastackvec}{\textbf{\textrm{s}}}
\newcommand{\modelstackvec}{\textbf{\textrm{m}}}
\newcommand{\boxsizeforgrad}{6^{\prime} \times 6^{\prime}}

\newcommand{\howmanyclustersinlownoisesims}{10,000}

\newcommand{\detectionormeasurement}{detection}
\newcommand{\degree}{\ensuremath{^\circ}}

\newcommand{\mvir}{M$_{200m}$}

\newcommand{\kappaonehalomz}{$\kappa^{1h}(\rm{M},z)$}
\newcommand{\kappatwohalomz}{$\kappa^{2h}(\rm{M}, z)$}
\newcommand{\kappatotalmz}{$\kappa^{tot}(\rm{M}, z)$}

\newcommand{\boxsize}{$200^{\prime} \times 200^{\prime}$}
\newcommand{\pixres}{$1^{\prime}$}
\newcommand{\MLboxsize}{$10' \times 10'$}
\newcommand{\whichyear}{\mbox{Year-3}}
\newcommand{\whichsample}{full}
\newcommand{\whichcatversion}{\texttt{y3\_gold:v6.4.22+2}}

\newcommand{\ML}{\mbox{${\rm M}-\lambda$} }

\newcommand{\ukam}{\ensuremath{\mu}{\rm K{\text -}arcmin}}
\newcommand{\sqdeg}{deg$^{2}$}

\newcommand{\desrm}{DES redMaPPer}
\newcommand{\RM}{redMaPPer} 
\newcommand{\msol}{\ensuremath{\mbox{M}_{\odot}}}
\newcommand{\munits}{\times 10^{14}~\msol}

\newcommand{\comment}[1]{}
\begin{document}

\ifdefined\PRformat
\begin{flushright}
\begin{tabular}{r}
 \footnotesize FERMILAB-PUB-19-429-AE \\
\footnotesize DES-2018-0395
\end{tabular}
\end{flushright}
\fi

\title{A Detection of CMB-Cluster Lensing using Polarization Data from \sptpol}
\author{S.~Raghunathan} 
\email{sri@physics.ucla.edu}
\affiliation{Department of Physics and Astronomy, University of California, Los Angeles, CA, USA 90095} \affiliation{School of Physics, University of Melbourne, Parkville, VIC 3010, Australia}

\author{S.~Patil} \affiliation{School of Physics, University of Melbourne, Parkville, VIC 3010, Australia}

\author{E.~Baxter} \affiliation{Department of Physics and Astronomy, University of Pennsylvania, Philadelphia, PA 19104, USA}

\author{B.~A.~Benson} \affiliation{Fermi National Accelerator Laboratory, MS209, P.O. Box 500, Batavia, IL 60510} \affiliation{Kavli Institute for Cosmological Physics, University of Chicago, 5640 South Ellis Avenue, Chicago, IL, USA 60637} \affiliation{Department of Astronomy and Astrophysics, University of Chicago, 5640 South Ellis Avenue, Chicago, IL, USA 60637}

\author{L.~E.~Bleem} \affiliation{High Energy Physics Division, Argonne National Laboratory, 9700 S. Cass Avenue, Argonne, IL, USA 60439} \affiliation{Kavli Institute for Cosmological Physics, University of Chicago, 5640 South Ellis Avenue, Chicago, IL, USA 60637}

\author{T.~M.~Crawford} \affiliation{Kavli Institute for Cosmological Physics, University of Chicago, 5640 South Ellis Avenue, Chicago, IL, USA 60637} \affiliation{Department of Astronomy and Astrophysics, University of Chicago, 5640 South Ellis Avenue, Chicago, IL, USA 60637}

\author{G.~P.~Holder} \affiliation{Astronomy Department, University of Illinois at Urbana-Champaign, 1002 W. Green Street, Urbana, IL 61801, USA} \affiliation{Department of Physics, University of Illinois Urbana-Champaign, 1110 W. Green Street, Urbana, IL 61801, USA}  \affiliation{Canadian Institute for Advanced Research, CIFAR Program in Gravity and the Extreme Universe, Toronto, ON, M5G 1Z8, Canada}

\author{T.~McClintock} \affiliation{Department of Physics, University of Arizona, Tucson, AZ 85721, USA}

\author{C.~L.~Reichardt} \affiliation{School of Physics, University of Melbourne, Parkville, VIC 3010, Australia}

\author{T.~N.~Varga} \affiliation{Max Planck Institute for Extraterrestrial Physics, Giessenbachstrasse, 85748 Garching, Germany} \affiliation{Universit\"ats-Sternwarte, Fakult\"at f\"ur Physik, LudwigMaximilians Universit\"at M\"unchen, Scheinerstr. 1, 81679 M\"unchen, Germany}

% ...



\newcommand{\abstracttext}
{
We report the first \detectionormeasurement{} of gravitational lensing due to galaxy clusters using only the polarization of the cosmic microwave background (CMB).
The lensing signal is obtained using a new estimator that extracts the lensing dipole signature from stacked images formed by rotating the cluster-centered Stokes $Q/U$ map cutouts along the direction of the locally measured background CMB polarization gradient. 
Using data from the SPTpol 500 deg$^{2}$ survey at the locations of roughly 18,000 clusters with richness $\lambda \ge 10$ from the Dark Energy Survey (DES) \whichyear{} \whichsample{} galaxy cluster catalog, we detect lensing at \howmanysigmastackedforpolaboverichten$\sigma$.  
The mean stacked mass of the selected sample is found to be \desrmfullsamplemeanmasspolsysaboverichten which is in good agreement with optical weak lensing based estimates using DES data and CMB-lensing based estimates using SPTpol temperature data.
This measurement is a key first step for cluster cosmology with future low-noise CMB surveys, like CMB-S4, for which CMB polarization will be the primary channel for cluster lensing measurements.
}

\date{Accepted XXX. Received YYY; in original form ZZZ}
\begin{abstract}
\abstracttext
\end{abstract}

\ifdefined\PRformat
\maketitle
\fi


\emph{Introduction. ---} 
Galaxy clusters are the most massive gravitationally bound structures in the Universe.
Measuring their abundance as a function of mass and redshift can provide tight constraints on the cosmological parameters that influence the geometry and growth of structures in the Universe \citep[see][for a review]{allen11} that are complementary to baryon acoustic oscillations (BAO) or cosmic microwave background (CMB) datasets.
The independent measurements of cluster abundance, BAO, and CMB, which have different parameter degeneracies, can be combined to obtain even stronger constraints \citep{wang05, mantz08, vikhlinin09, mantz10c, rozo10, hasselfield13, mantz15, placksz15, dehaan16, salvati17, bocquet19}.
However, the cluster abundance measurements rely on precise mass measurements, which are currently limited by uncertainties in the conversion of the survey observable to cluster mass \citep{linden14}. 
Upcoming large surveys are forecasted to detect tens of thousands of galaxy clusters, an order of magnitude more than current surveys \citep{lsst09, erosita12, cmbs4-sb1}.
Of these, CMB surveys, in which galaxy clusters are observed via redshift-independent Sunyaev-Zel'dovich (SZ) effect, will return $\gtrsim 10,000$ clusters above $z \ge 1$ \citep{cmbs4-sb1}. 
Given such an enormous increase in the sample size compared to the current surveys, it is crucial to develop robust methods to measure cluster masses accurately. 

In contrast to other cluster observables (optical richness, SZ flux, and X-ray flux), gravitational lensing of galaxies or the CMB offers an unbiased mass measurement since lensing exactly traces the underlying matter distribution. 
Weak lensing measurements of galaxies have high signal-to-noise (\snr) at low redshifts, but the \snr{} falls steeply at high redshifts with the number of distant lensed background galaxies observed with sufficiently high \snr{} to facilitate lensing.

By contrast, since the CMB originates behind all of the clusters, lensing of the CMB by clusters is a highly promising tool for measuring masses of clusters above $z \ge 1$ \citep{lewis06a}.
The CMB-cluster signal can be observed with both temperature and polarization anisotropies of the CMB.
As the amplitude of the lensing signal is proportional to the local CMB gradient, lensing of the brighter CMB temperature anisotropies yields a higher \snr{} compared to polarization. 
A number of experiments have now detected the CMB-cluster lensing signal in temperature \citep{madhavacheril15, baxter15, placksz15, geach17, baxter18, raghunathan17b, raghunathan18}, yielding mass constraints at the 10\% level \citep{geach17}. 
However, CMB temperature data are susceptible to foregrounds that set an effective noise floor for future measurements.
CMB polarization, on the other hand, is robust to foregrounds as contaminating signals from the galaxy cluster itself and other foregrounds are much lower in polarization than temperature \citep[see Fig. 2 of][]{raghunathan17a}. 
As a result, polarized CMB-cluster lensing will be crucial to the cluster mass constraints from next generation low-noise surveys \citep{raghunathan17a}.

Several polarized CMB-cluster lensing estimators have been proposed \citep{lewis06a, hu07, yoo10}, however none have yet been demonstrated on data. 
In this work we detect, for the first time, the CMB-cluster lensing signal from polarization data alone. 
We develop a new estimator that extracts the lensing dipole signature from the CMB maps by rotating the cluster-centered cutouts along the direction of the local background CMB polarization gradient.
The method is easy to implement and computationally much less expensive compared to the traditional maximum likelihood estimator \citep{dodelson04, lewis06a, baxter15, raghunathan17a} which models the lensing signal using a large suite of simulations.
We apply this estimator to the SPTpol 500 \sqdeg{} polarization Stokes $Q/U$ maps at the location of clusters from the Dark Energy Survey (DES) \whichyear{} catalog. 
We reject the null hypothesis of no lensing at $\howmanysigmastackedforpolaboverichten \sigma$ in the combined Q/U maps.
This result demonstrates the viability of achieving sub-percent level mass constraints \citep{raghunathan17a} from next-generation CMB surveys like CMB-S4 \citep{cmbs4-sb1}.

Throughout this work, we use the \planck{} 2015 best-fit $\Lambda$CDM cosmology 
\citep{planck15-13} with $h=0.67$, and assume the absence of primordial B-modes.
The lensed CMB power spectra were obtained using \texttt{CAMB} \citep{lewis00}. 
All the halo quantities are defined with respect to a sphere within which the average mass density is 200 times the mean density of the Universe at the halo redshift.  


\emph{Dataset I: The SPTpol 500 deg$^{2}$ survey. ---}
We use two datasets in this work. 
The first is the 150\,GHz Stokes \QU{} polarization maps of a 500\,deg$^{2}$ region (R.A.~=~22h to 2h;  Decl. = \mbox{-65\degree} to -50\degree) from the SPTpol survey. 
The South Pole Telescope (SPT) is a \mbox{10-m} telescope located at the Amundsen-Scott South Pole station \citep{padin08, carlstrom11} and 
\sptpol{} was the second camera on the SPT. 
It has 1176 polarization-sensitive transition-edge-sensor bolometers \citep{austermann12} and roughly a $1.^{\prime}2$ FWHM beam at 150\,GHz.
The white noise level of the polarization maps is $\Delta_{P} \sim 7\,\ukam$. 
The maps used in this analysis were made in the Sanson-Flamsteed flat-sky projection with a pixel resolution of \pixres. 
From these Stokes \QU{} maps, we remove an estimate of the temperature-to-polarization leakage ($T\rightarrow P$) as $X = X - \epsilon_{X} T$ where $X \in [Q,U]$, $\epsilon_{Q} = 1.65\%$, and $\epsilon_{U} = 0.71\%$.
Unaccounted for, $T\rightarrow P$ would introduce temperature signal from the galaxy clusters, such as the SZ effects \citep{sunyaev72, sunyaev80b} or emission from radio galaxies and dusty galaxies, into the polarization maps. 
More details about the map making procedure can be found in \citet{henning18}. 

\emph{Dataset II: DES cluster catalog. ---}
The second data product used in the analysis is a sample of optically selected clusters from the DES, which is an optical to near-infrared survey from the Atacama region in northern Chile. 
In this work, we use a cluster catalog selected by the \RM{} (RM) algorithm \citep{rykoff14} using DES \whichyear{} observations of $\sim 3000$ \sqdeg{}, specifically we use the \whichsample{} flux-limited catalog version: \whichcatversion{}.  
We select all clusters with richness $\lambda \geq 10$ within the SPTpol survey area, where we exclude any cluster within 30$^\prime$  of the survey boundary or within 10$^\prime$ of a source with $S_{\rm 150 GHz} > 6.4$ mJy.
In total we work with \howmanyclustersaboverichten{} clusters, of which \howmanyclustersaboverichtwenty{} have richness $\lambda \ge 20$. 
The cluster redshifts are estimated photometrically with uncertainties of  $\hat{\sigma}_{z} = 0.01 (1+z)$ \citep{rozo15}. 
We neglect redshift uncertainties in this work since the impact of photo$-z$ errors on CMB-lensing masses is negligible \citep{raghunathan17a}.  
The redshifts span $0.1 \le z \le 0.95$ with a median value of $z_{\rm med} = 0.72$.

The low-richness ($\lambda < 20$) haloes are included to improve the lensing \snr{} as the goal here is only to make the first measurement of the polarized CMB-cluster lensing signal. 
Since these low mass objects are not well characterized by the RM algorithm, we caution the reader when using results from the low-richness objects in this work for any cosmological analysis. 



\emph{Lensing estimator. ---}
On scales corresponding to the angular size of a galaxy cluster, the primordial CMB is exponentially damped \cite{silk68} and the field can be well approximated by a gradient. 
When a galaxy cluster lenses this CMB gradient field, it produces a dipole-like pattern \cite{seljak00, lewis06a} that is oriented along the direction of the gradient \cite[see Fig. 1 of][]{lewis06a}. 
This is the basis for the lensing estimator developed here which uses the following steps to extract the lensing dipole and constrain the cluster masses: 
\begin{enumerate}
\item[1.]{Extract \MLboxsize{} ${\rm N_{clus}}$ cluster-centered or $\rm{N_{rand}}$ random cutouts $\tilde{\datavec}$ from the Stokes \QU{} maps.}
\item[2.]{Determine the median value of the gradient direction $\theta_{\nabla} = {\rm tan}^{-1} (\nabla_{y}/\nabla_{x})$ in every \QU{} cutout.}
\item[3.]{Rotate $i^{th}$ cluster cutout $\tilde{\datavec}_{i}$ along $\theta_{\nabla, i}$ to obtain $\datavec_{i}$.}
\item[4.]{Determine weights $w$ (see below) for each cutout and stack the mean-subtracted cutouts to obtain the weighted stacked signal $\datastackvec_{c}$ ($\datastackvec_{r}$) at the cluster (random) locations.}
\item[5.]{Obtain the final lensing dipole signal as: $\datastackvec =  \datastackvec_{c} - \datastackvec_{r}$.}
\end{enumerate}

The gradient direction determination in step 2 is limited to a $\boxsizeforgrad$ region in each cutout and to reduce the noise penalty in the gradient estimation, we apply a Wiener filter of the form
\begin{eqnarray}
W_{\ell} &=&   \left\{
\begin{array}{l l}
C_{\ell} (C_{\ell} + N_{\ell})^{-1}&, ~\ell \le 2000\\
0&, ~{\rm otherwise}
\end{array}\right.
\label{eq_grad_filter}
\end{eqnarray} where $N_{\ell}$ is the noise spectrum and $C_{\ell}$ corresponds to 
$C_{\ell}^{QQ}, C_{\ell}^{UU}$ calculated from $C_{\ell}^{EE}, C_{\ell}^{BB}$. 
Note that we use Eq.(\ref{eq_grad_filter}) only for the gradient angle determination and the stack is obtained from the unfiltered, rotated \MLboxsize{} cutouts. 
We observe no significant change in our results when we replace $N_{\ell}$ in Eq.(\ref{eq_grad_filter}) by the full 2D noise power spectral density.

The weight \mbox{$w_{i} = w_{i, n} w_{i, g}$} assigned to cluster $i$ while stacking in step 4 can be decomposed into two pieces: one based on the inverse noise variance $\sigma_{i}^{2}$ at the location $i$; and the other using the median value of the magnitude of the local gradient $\sqrt{\nabla_{y_{i}}^{2} + \nabla_{x_{i}}^{2}}$.  
The latter serves to improve the \snr{} since the lensing amplitude is proportional to the gradient amplitude.

The stack $\datastackvec_{c}$ from cluster locations, however, is dominated by the mean large-scale CMB polarization gradient that we call the background.
We estimate and subtract the background $\datastackvec_{r}$ from a similar set of operations on $\rm{N_{rand}} = 50,000$ random locations.
The final rotated, background subtracted signal stack is constructed as 
 \begin{equation}
\datastackvec \equiv \datastackvec_{c} - \datastackvec_{r} = \frac{\sum_{c}^{{\rm N_{clus}}}  w_c \left[ \datavec_{c} - \left< \datavec_{c} \right> \right]}{ \sum_{c}^{{\rm N_{clus}}} w_c} - \frac{ \sum_{r}^{\rm{N_{rand}}} w_r \left[ \datavec_{r} - \left< \datavec_{r} \right> \right]}{ \sum_{r}^{\rm{N_{rand}}} w_r}
 \end{equation} 
where $\datavec$ represents the \QU{} cutout at a cluster location $c$ or a random location $r$.
Along with the lensing dipole, $\datastackvec$ includes contribution from other sources: foregrounds, instrumental noise, and the residual large-scale CMB gradient. 

\begin{figure}
\centering
\includegraphics[width=0.44\textwidth, keepaspectratio]{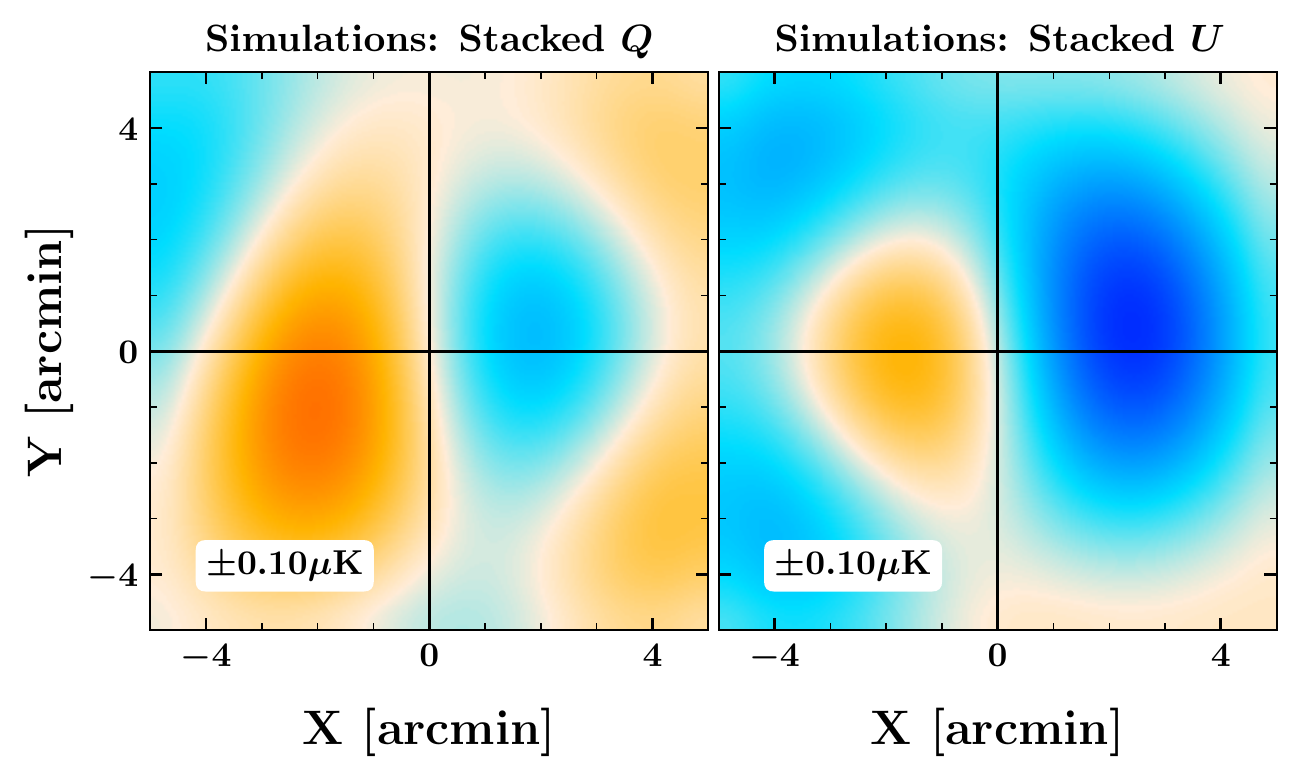}
\caption{Example lensing dipole signal extracted from low-noise simulated \QU{} stacks. The stack includes contributions from \howmanyclustersinlownoisesims{} clusters. The background, estimated from random locations, has been subtracted to remove the large-scale CMB gradient signals from both the panels.}
\label{fig_lensing_dipole_sims_lownoise}
\end{figure}

For visualization purposes, in Fig.~\ref{fig_lensing_dipole_sims_lownoise} we show the recovered lensing dipole signal \QU{} stack for low-noise ($\Delta_{P} = 0.1\ \ukam$) simulations.
The stack contains signal from ${\rm N_{clus}} = \howmanyclustersinlownoisesims$ clusters with (\mvir,\ z) fixed at ($2\munits,\ 0.7$). 
The presence of the dipole signal in the stacked \QU{} maps is the evidence for lensing. 
In the absence of lensing, the stacks will be consistent with null signals. 

Using the signal stack $\datastackvec$, we build a likelihood function
\begin{equation}
-2\ {\rm ln}\mathcal{L}({\rm M}|\datastackvec) =  \sum_{\rm pixels} \left( \datastackvec - \modelstackvec \right) \hat{{\bf C}}^{-1}  \left( \datastackvec - \modelstackvec \right)^{T}.
\label{eq_MLE_likelihood}
\end{equation}
where $\modelstackvec$ represents the model \ifdefined\PRformat\else{(see next section)}\fi
and the covariance matrix $\hat{{\bf C}}$ is estimated using a jackknife re-sampling technique
by dividing the survey region into $N$ sub-fields 
\begin{equation}
\hat{{\bf C}} = \frac{N-1}{N} \sum\limits_{j=1}^{N} \left[\datastackvec_{j} - \left<\datastackvec\right>\right] \left[\datastackvec_{j} - \left<\datastackvec\right>\right]^{T},
\label{eq_JK_cov}
\end{equation}
where $\datastackvec_{j}$ is the stack of all the clusters in the $j^{th}$ sub-field and  $\left<\datastackvec\right>$ is the ensemble average of all the sub-fields.
$\hat{{\bf C}}$ properly captures all sources of noise since it is estimated from the data itself.

\emph{Lensing dipole models. ---}
For Eq.(\ref{eq_MLE_likelihood}) we construct a model stack, $\modelstackvec \equiv \modelstackvec({\rm M})$, using the above steps, except at step 1 we replace the data vector, $\datavec$, with no-noise cluster-lensed simulations described below. 

For each mass, ${\rm M}$, in the parameter grid we generate ${\rm N_{clus}}$ cluster-lensed realizations of the Stokes \QU{} maps. 
This is done by generating convergence profiles at each of the measured DES cluster redshifts for each mass.
We follow steps 2-4 to obtain the stacked model $\modelstackvec_{c}({\rm M})$. 
The mean background gradient CMB in this case simply corresponds to $\modelstackvec_{r} \equiv \modelstackvec_{c}({\rm M}=0)$ and we remove that from models calculated at all the other masses in the parameter grid. 
We use a flat prior for mass in the range ${\rm M} \in [0,4] \munits$ and divide the parameter grid linearly in bins $\Delta {\rm M} = 0.1 \munits$.
From the likelihood, we measure the median mass and $1\sigma$ uncertainty, defined by the 16 to 84 percent confidence range.

Note that the uncertainties $\delta \theta_{\nabla}$ in step 2 will be lower in no-noise models compared to the data. 
These errors lead to suboptimal stacking of the lensing dipole and will result in a bias towards low mass if not accounted for in the model. 
Subsequently, we add noise in the simulations similar to that of the data \ifdefined\PRformat\else{(see \S\ref{sec_sptpol})}\fi only when determining $\theta_{\nabla}$. 
This ensures that the uncertainties $\delta \theta_{\nabla}$ caused by instrumental noise in the data are also replicated in the models. 

\emph{Simulations. ---}
The simulations used to create the lensing dipoles and mock datasets follow our previous work \cite{raghunathan17a}. 
Briefly, the Stokes \QU{} simulations are created from Gaussian realizations of the CMB $E$- and $B$-mode maps using flat-sky approximations and span \boxsize.
The convergence profile used to lens the $E$- and $B$-mode maps includes contributions from \kappatotalmz = \kappaonehalomz + \kappatwohalomz. 
We use Navarro-Frenk-White (NFW) \cite{navarro96} profile to model the one-halo term \kappaonehalomz{} \cite{bartelmann96} and follow the prescription given in \citet{oguri11} for the lensing contribution from correlated structures \kappatwohalomz{} \cite{seljak00a, cooray02}. 
We also correct \kappaonehalomz{} to account for uncertainties in the cluster centroids as \cite{oguri10}
\begin{eqnarray}
\tilde{\kappa}(\ell) = \kappa(\ell) \left[ (1-f_{\rm mis}) + f_{\rm mis}\ {\rm exp}\left( -\frac{1}{2} \sigma^{2}_{s} \ell^{2} \right) \right]
\end{eqnarray}
We set the fraction of mis-centered clusters to $f_{\rm mis} = 0.22$ \cite{rykoff16} and $\sigma_{\rm s} = \sigma_{\rm R}/D_{A}(z)$. 
The amount of mis-centering $\sigma_{\rm R}$,  which is a fraction of the cluster radius ($R_{\lambda} = (\lambda/100)^{0.2} h^{-1}$ Mpc) is modeled as a Rayleigh distribution with $\sigma_{\rm R}  = c_{\rm mis} R_{\lambda}$ where ${\rm ln}\ c_{\rm mis} = -1.13 \pm 0.22$ \cite{rykoff16}. 
$D_{A}(z)$ in the above equation is the angular diameter distance at the cluster redshift $z$. 

We smooth the \QU{} maps using the measured beam function for SPTpol \citep{henning18} and account for the information lost during the map-making process due to the filtering applied to the data.  
We approximate the filtering as a 2D transfer function \citep{baxter18, raghunathan18} given as $F_{\bar{\ell}} = e^{-(\ell_{1}/\ell_{x})^{6}} e^{-(\ell_{x}/\ell_{2})^{6}}$ with $\ell_{1}$ = 300, and $\ell_{2}$ = 20,000. 
The two terms can be understood as high-pass and low-pass filters in the scan direction respectively. 
To generate mock datasets for pipeline validation, we also add Gaussian realizations of the instrumental noise at the desired level.
The central \MLboxsize{} cutouts are extracted from the simulated maps and passed through the rest of the pipeline steps described earlier \ifdefined\PRformat\else{ in \S\ref{sec_lensing_estimator_details}}\fi to obtain the model or the mock datasets for the pipeline validation.



\emph{Pipeline validation. ---}
We now validate the lensing pipeline and estimate the expected lensing \snr{} for the DES clusters. 
To the lensed simulated \QU{} maps we add instrumental noise using the noise power $N_{\ell}$ measured from the SPTpol \QU{} maps. 
The number of simulated clusters and their redshifts and richnesses match the real values in the \desrm{} \whichyear{} \whichsample{} sample. 
The richnesses and redshifts are converted to cluster masses using the \ML{} relation: ${\rm M} = A \left(\frac{\lambda}{30}\right)^{\alpha} \left(\frac{1+z}{1+0.5}\right)^{\beta}$
where $A$ is a normalization, and the exponents $\alpha$ and $\beta$ are richness and redshift evolution parameters, respectively. 
We use the best-fit values for these parameters obtained from DES weak-lensing analysis \citep{mcclintock18}, namely \mbox{$A = 3.08\ \munits$}, $\alpha = 1.36$, and $\beta = -0.3$.
The mean mass of the simulated sample is \mbox{\mvir = \desrmfullsamplemeanmassabovetennoerrors}. 
We note that the DES \ML{} relation has been calibrated only using clusters with $\lambda \ge 20$
and the relation cannot be fully trusted for lower richness objects. 
However, we employ the relation here only to obtain a rough estimate of the final lensing \snr.

\begin{figure}
\centering
\includegraphics[width=0.45\textwidth, keepaspectratio]{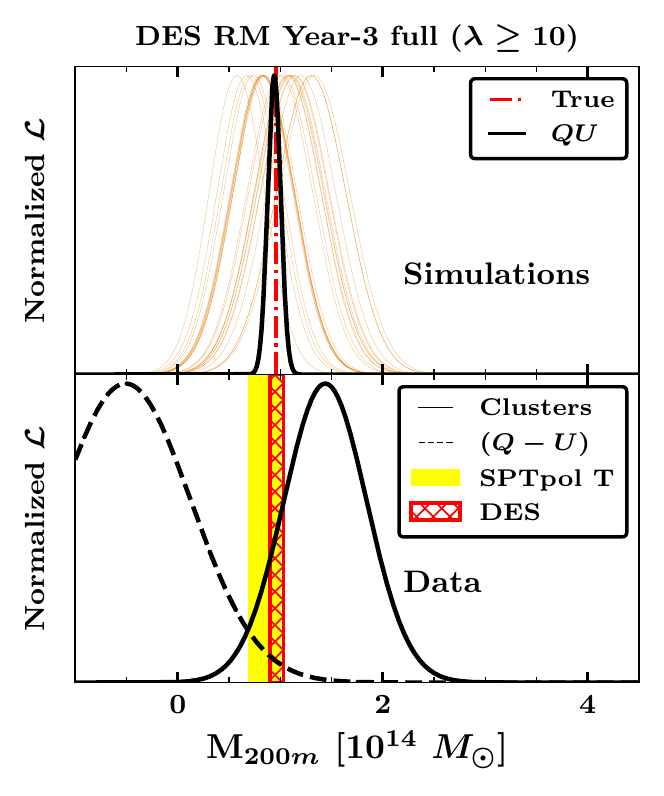}
\caption{
Lensing mass constraints of DES RM clusters using polarization-only data from the \sptpol{} survey at the location of \howmanyclustersaboverichten{} clusters. 
In the top panel, the light shaded curves are for 25 individual simulations and their combined likelihood is the thicker solid curve. 
The true mass from DES weak lensing  measurements is given as the red dash-dotted line.
The result from stacked SPTpol data (bottom panel) is in good agreement with the weak lensing measurements from DES (red region) and the SPTpol temperature result (yellow region).
The ($Q-U$) null test is shown as the dashed curve in the bottom panel. 
}
\label{fig_MLE_results}
\end{figure}

Next we extract the lensing dipole from the simulated maps by following the steps 1-5 described in the methods section. 
We combine the data from \QU{} into a single $QU$ map vector. 
The covariance in this case $\hat{{\bf C}} \equiv \hat{{\bf C}}_{QU}$ also includes the covariance between the $Q$ and $U$ cutouts. 
The results for this $QU$ estimator are presented in the top panel of Fig.~\ref{fig_MLE_results}. 
Each light shaded curve represents one simulation run for the DES cluster sample. 
The combined result from 25 runs, \mvir = \desrmfullsamplemeanmassabovetensims, plotted as the thicker black curve, is within $0.25\sigma$ of the input mass (red dash-dotted line).
We evaluate the likelihood of the null hypothesis of no lensing using the statistic, 
\snr{} = $\sqrt{\Delta \chi^2} = \sqrt{2\left[ {\rm ln}\mathcal{L}(M_{200m} = M_{\rm fit}) - {\rm ln}\mathcal{L}(M_{200m} = 0) \right]}$ 
and obtain an average lensing \snr{} of $4.3\sigma$ from these simulations translating to roughly 25\% constraints in the stacked cluster mass.

\emph{Systematics. ---}
Systematics in our measurement arise from the following sources: (a) assumption of a background cosmology for model generation; (b) incorrect cluster profile; and (c) the uncertainties in the DES mis-centering model. 
The biases are quantified using the mock datasets for $10\times$ more clusters, but after including the modifications described below.
In all these cases, the models remain fixed to the fiducial \planck{} 2015 cosmology and the standard NFW profiles. 

We quantify the bias due to the mis-match between the underlying and the assumed cosmology by re-running the simulations using a different $C_{\ell}$ within the $1\sigma$ errors of the cosmological parameters obtained by \planck{} (ignoring the correlations between the parameters). 
This change modifies the power in \QU{} and also the lensing convergence profiles. 
To quantify the errors due to the assumption of a NFW profile for DES clusters, we replace the NFW profile in the mock dataset generation with an Einasto profile \cite{einasto89}. 
Finally, to assess the effect of uncertainties in mis-centering, we create a new mis-centering distribution by increasing the values of $f_{\rm mis}$ and ${\rm ln}\ c_{\rm mis}$ by their 1$\sigma$ uncertainties and use the result to calculate the smeared convergence $\kappa^{\prime}_{1h}$.

In all cases the shifts in the inferred lensing mass are negligible compared to the $25\%$ constraints on the masses that we expect.  
Specifically we obtain the following biases: 1.5\% ($0.15\sigma$), 0.5\% ($<0.1\sigma$), and 1.1\% ($0.12\sigma$) for the three cases with a combined error budget of 2\% ($0.22\sigma$) for a sample that contains $10\times$ more clusters.
Given that the sample size in this work is much smaller than for the tests considered here, we expect the effects of systematics to be minimal and our results to be dominated by statistical errors.

\emph{Polarization lensing measurement. ---}
In this analysis, we constrain the mass of a sample of clusters selected from the DES \whichyear{} data set using the RM algorithm. 
The lensing masses for two samples, $\lambda \ge 10$ and $\lambda \ge 20$, are given in Table \ref{tab_lensing_mass}. 
The table also contains the comparisons to the weak-lensing measurements from DES \citep{mcclintock18} and SPTpol temperature results \citep{raghunathan18}  by converting the richness estimates into mass using the \ML scaling relation reported in those works. 
The posterior distribution for the weighted mean of the cluster masses is shown as the black solid curve in the bottom panel of Fig.~\ref{fig_MLE_results}. 
The recovered cluster mass from polarization is within $1.3-1.5\sigma$ of both the results.
Note that the contribution from \kappatwohalomz{} is included in the model here.
Ignoring the \kappatwohalomz{} term moves the lensing mass higher, as expected, by 9\%. 

As a further systematics test, we test whether results are dominated by either $Q$ or $U$ by obtaining mass estimates from $Q$ and $U$ separately.
We obtain \desrmfullsamplemeanmasspolsysaboverichtenforQ{} and \desrmfullsamplemeanmasspolsysaboverichtenforU{} for $Q$ and $U$ respectively for the $\lambda \ge 10$ sample. 
Furthermore, we perform a null test with by differencing the signals from $Q$ and $U$, to check if it is consistent with random fluctuations.
The lensing mass of \desrmfullsamplemeanmasspolsysaboverichtenforQminusUnulltest{} shown as the dashed curve in the bottom panel of Fig.~\ref{fig_MLE_results} confirms the null signal. 
Another test performed by stacking \howmanyclustersaboverichtenfornulltest{} random locations, also returns a lensing mass of \desrmfullsamplemeanmasspolsysaboverichtenulltest, consistent with \mvir = 0. 

\begin{table}
\caption{Recovered lensing masses of the DES RM cluster sample.}
\centering
\begin{tabular}{|p{2cm} | p{2cm}@{}| p{2cm} |  p{2cm} |}
\hline
\multirow{2}{*}{Sample} & \multicolumn{3}{c|}{Lensing mass \mvir $\munits$}\\
\cline{2-4}
& This work & DES & SPTpol-T \\\hline
$\lambda \ge 10$ & \desrmfullsamplemeanmasspolsysaboverichtennounits & \desrmfullsamplemeanmassaboveten & \desrmfullsamplemeanmasstempaboveten \\\hline
$\lambda \ge 20$ & \desrmfullsamplemeanmasspolsysaboverichtwentynounits & \desrmfullsamplemeanmassabovetwenty & \desrmfullsamplemeanmasstempabovetwenty \\\hline
\end{tabular}
\label{tab_lensing_mass}
\end{table}

\begin{figure}
\centering
\includegraphics[width=0.45\textwidth, keepaspectratio]{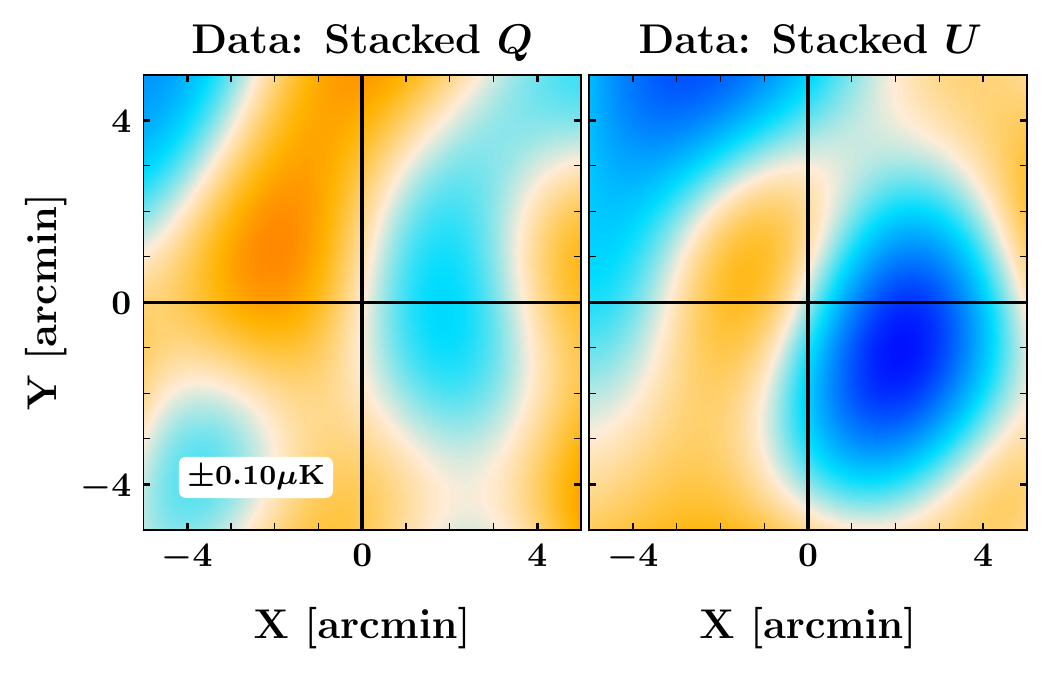}
\caption{
Rotated, background-subtracted $Q$ and $U$ stacks from the SPTpol data showing the cluster lensing dipole signals.
Unlike in Fig.~\ref{fig_lensing_dipole_sims_lownoise}, these images have been filtered to remove the small-scale noise for illustrative purposes. 
}
\label{fig_lensing_dipole_data}
\end{figure}

For visual illustration, the rotated cluster stacks are presented in Fig.~\ref{fig_lensing_dipole_data}. 
Since the noise levels of the SPTpol  maps are much higher than in Fig.~\ref{fig_lensing_dipole_sims_lownoise}, we apply additional filtering to remove the small-scale noise in the figure. 
We adopt a Wiener filter similar to Eq.(\ref{eq_grad_filter}) but after replacing $C_{\ell}$ by the power spectra of the \QU{} lensing dipole signal corresponding to the lensing mass obtained above, scaling $N_{\ell}$ by $\sqrt{\rm N_{clus}}$ in the stack, and low-pass filtering the stack below $\ell \le 4000$. 
This filter is not used in the actual analysis.

Finally, we find that the no-lensing hypothesis is disfavored at \howmanysigmastackedforpolaboverichten$\sigma$ (\howmanysigmastackedforpolaboverichtwenty$\sigma$) for the $\lambda \ge 10$ ( $\lambda \ge 20$) sample which is in good agreement with the expectations from simulations. 
This represents the first \detectionormeasurement{} of the CMB-cluster lensing signal in polarization data. 

\emph{Future prospects. ---}
The estimator developed in this work can also be applied to temperature data. 
When using the temperature data, however, we must additionally fit for the rotationally invariant thermal SZ signal in the stacked cutouts and other possible sources of cluster correlated foregrounds.  
Similarly, the performance of the estimator must be compared to other lensing estimators \citep{hu07, yoo10, raghunathan17a} to determine the optimal method of CMB-cluster lensing reconstruction both in terms of the computational requirements and the sensitivity. 
We defer a detailed investigation of these to a future work. 

For future experiments, CMB polarization-based results will be increasingly important for CMB-lensing based cluster mass estimates.
The systematics introduced by astrophysical foregrounds, which are largely unpolarized, is much reduced in CMB polarization compared to temperature.
For example, sources in CMB maps have been measured to have a fractional polarization of $\sim 3\%$ with random polarization angles \citep[recently, ][]{datta18, gupta19}.
In \citet{raghunathan17a}, we showed that polarized point sources cause negligible bias in CMB-cluster lensing even at polarization fractions higher than this. 
The polarization of the SZ effect should also have negligible impact, and is expected to be two orders of magnitude smaller \citep{carlstrom02, hall14, yasini16} than the lensing signal expected from the clusters. 

This measurement is the first step towards achieving precise mass constraints \citep{raghunathan17a} from next-generation CMB surveys like CMB-S4 \citep{cmbs4-sb1} and SPT-3G \citep{bender18}, and will be important to maximize the cosmological constraining power of future cluster surveys.

\emph{Acknowledgments. ---}
The authors thank Andrew Ludwig, Nickolas McColl, Siavash Yasini, and the three anonymous reviewers for their valuable feedback on the manuscript. 

SR acknowledges partial support from the Laby Foundation. 
SP acknowledges support from Melbourne International Engagement Award (MIPP) and Laby Travel Bursary.
The UCLA authors acknowledge support from NSF grants AST-1716965 and CSSI-1835865. 
Melbourne group acknowledges support from the Australian Research Council's Discovery Projects scheme (DP150103208). 
LB's work was supported under the U.S. Department of Energy contract DE-AC02-06CH11357.
We acknowledge the use of \texttt{CAMB} \citep{lewis00} software. 

This work was performed in the context of the South Pole Telescope scientific program. SPT is supported by the National Science Foundation through grant PLR-1248097.  Partial support is also provided by the NSF Physics Frontier Center grant PHY-1125897 to the Kavli Institute of Cosmological Physics at the University of Chicago, the Kavli Foundation and the Gordon and Betty Moore Foundation grant GBMF 947. This research used resources of the National Energy Research Scientific Computing Center (NERSC), a DOE Office of Science User Facility supported by the Office of Science of the U.S. Department of Energy under Contract No. DE-AC02-05CH11231.

Funding for the DES Projects has been provided by the U.S. Department of Energy, the U.S. National Science Foundation, the Ministry of Science and Education of Spain, 
the Science and Technology Facilities Council of the United Kingdom, the Higher Education Funding Council for England, the National Center for Supercomputing 
Applications at the University of Illinois at Urbana-Champaign, the Kavli Institute of Cosmological Physics at the University of Chicago, 
the Center for Cosmology and Astro-Particle Physics at the Ohio State University,
the Mitchell Institute for Fundamental Physics and Astronomy at Texas A\&M University, Financiadora de Estudos e Projetos, 
Funda{\c c}{\~a}o Carlos Chagas Filho de Amparo {\`a} Pesquisa do Estado do Rio de Janeiro, Conselho Nacional de Desenvolvimento Cient{\'i}fico e Tecnol{\'o}gico and 
the Minist{\'e}rio da Ci{\^e}ncia, Tecnologia e Inova{\c c}{\~a}o, the Deutsche Forschungsgemeinschaft and the Collaborating Institutions in the Dark Energy Survey. 

The Collaborating Institutions are Argonne National Laboratory, the University of California at Santa Cruz, the University of Cambridge, Centro de Investigaciones Energ{\'e}ticas,

Medioambientales y Tecnol{\'o}gicas-Madrid, the University of Chicago, University College London, the DES-Brazil Consortium, the University of Edinburgh, 
the Eidgen{\"o}ssische Technische Hochschule (ETH) Z{\"u}rich, 
Fermi National Accelerator Laboratory, the University of Illinois at Urbana-Champaign, the Institut de Ci{\`e}ncies de l'Espai (IEEC/CSIC), 
the Institut de F{\'i}sica d'Altes Energies, Lawrence Berkeley National Laboratory, the Ludwig-Maximilians Universit{\"a}t M{\"u}nchen and the associated Excellence Cluster Universe, 
the University of Michigan, the National Optical Astronomy Observatory, the University of Nottingham, The Ohio State University, the University of Pennsylvania, the University of Portsmouth, 
SLAC National Accelerator Laboratory, Stanford University, the University of Sussex, Texas A\&M University, and the OzDES Membership Consortium.

Based in part on observations at Cerro Tololo Inter-American Observatory, National Optical Astronomy Observatory, which is operated by the Association of 
Universities for Research in Astronomy (AURA) under a cooperative agreement with the National Science Foundation.

The DES data management system is supported by the National Science Foundation under Grant Numbers AST-1138766 and AST-1536171.
The DES participants from Spanish institutions are partially supported by MINECO under grants AYA2015-71825, ESP2015-66861, FPA2015-68048, SEV-2016-0588, SEV-2016-0597, and MDM-2015-0509, 
some of which include ERDF funds from the European Union. IFAE is partially funded by the CERCA program of the Generalitat de Catalunya.
Research leading to these results has received funding from the European Research
Council under the European Union's Seventh Framework Program (FP7/2007-2013) including ERC grant agreements 240672, 291329, and 306478.
We  acknowledge support from the Australian Research Council Centre of Excellence for All-sky Astrophysics (CAASTRO), through project number CE110001020, and the Brazilian Instituto Nacional de Ci\^encia e Tecnologia (INCT) e-Universe (CNPq grant 465376/2014-2).

This manuscript has been authored by Fermi Research Alliance, LLC under Contract No. DE-AC02-07CH11359 with the U.S. Department of Energy, Office of Science, Office of High Energy Physics. The United States Government retains and the publisher, by accepting the article for publication, acknowledges that the United States Government retains a non-exclusive, paid-up, irrevocable, world-wide license to publish or reproduce the published form of this manuscript, or allow others to do so, for United States Government purposes.
This manuscript has been authored by Fermi Research Alliance, LLC under Contract No. DE-AC02-07CH11359 with the U.S. Department of Energy, Office of Science, Office of High Energy Physics. The United States Government retains and the publisher, by accepting the article for publication, acknowledges that the United States Government retains a non-exclusive, paid-up, irrevocable, world-wide license to publish or reproduce the published form of this manuscript, or allow others to do so, for United States Government purposes.


\bibliographystyle{apsrev4-1}
\bibliography{spt_extract}

\begin{thebibliography}{54}%
\makeatletter
\providecommand \@ifxundefined [1]{%
 \@ifx{#1\undefined}
}%
\providecommand \@ifnum [1]{%
 \ifnum #1\expandafter \@firstoftwo
 \else \expandafter \@secondoftwo
 \fi
}%
\providecommand \@ifx [1]{%
 \ifx #1\expandafter \@firstoftwo
 \else \expandafter \@secondoftwo
 \fi
}%
\providecommand \natexlab [1]{#1}%
\providecommand \enquote  [1]{``#1''}%
\providecommand \bibnamefont  [1]{#1}%
\providecommand \bibfnamefont [1]{#1}%
\providecommand \citenamefont [1]{#1}%
\providecommand \href@noop [0]{\@secondoftwo}%
\providecommand \href [0]{\begingroup \@sanitize@url \@href}%
\providecommand \@href[1]{\@@startlink{#1}\@@href}%
\providecommand \@@href[1]{\endgroup#1\@@endlink}%
\providecommand \@sanitize@url [0]{\catcode `\\12\catcode `\$12\catcode
  `\&12\catcode `\#12\catcode `\^12\catcode `\_12\catcode `\%12\relax}%
\providecommand \@@startlink[1]{}%
\providecommand \@@endlink[0]{}%
\providecommand \url  [0]{\begingroup\@sanitize@url \@url }%
\providecommand \@url [1]{\endgroup\@href {#1}{\urlprefix }}%
\providecommand \urlprefix  [0]{URL }%
\providecommand \Eprint [0]{\href }%
\providecommand \doibase [0]{http://dx.doi.org/}%
\providecommand \selectlanguage [0]{\@gobble}%
\providecommand \bibinfo  [0]{\@secondoftwo}%
\providecommand \bibfield  [0]{\@secondoftwo}%
\providecommand \translation [1]{[#1]}%
\providecommand \BibitemOpen [0]{}%
\providecommand \bibitemStop [0]{}%
\providecommand \bibitemNoStop [0]{.\EOS\space}%
\providecommand \EOS [0]{\spacefactor3000\relax}%
\providecommand \BibitemShut  [1]{\csname bibitem#1\endcsname}%
\let\auto@bib@innerbib\@empty
\bibitem [{\citenamefont {{Allen}}\ \emph {et~al.}(2011)\citenamefont
  {{Allen}}, \citenamefont {{Evrard}},\ and\ \citenamefont
  {{Mantz}}}]{allen11}%
  \BibitemOpen
  \bibfield  {author} {\bibinfo {author} {\bibfnamefont {S.~W.}\ \bibnamefont
  {{Allen}}}, \bibinfo {author} {\bibfnamefont {A.~E.}\ \bibnamefont
  {{Evrard}}}, \ and\ \bibinfo {author} {\bibfnamefont {A.~B.}\ \bibnamefont
  {{Mantz}}},\ }\href {\doibase 10.1146/annurev-astro-081710-102514} {\bibfield
   {journal} {\bibinfo  {journal} {\araa}\ }\textbf {\bibinfo {volume} {49}},\
  \bibinfo {pages} {409} (\bibinfo {year} {2011})},\ \Eprint
  {http://arxiv.org/abs/1103.4829} {arXiv:1103.4829 [astro-ph.CO]} \BibitemShut
  {NoStop}%
\bibitem [{\citenamefont {{Wang}}\ \emph {et~al.}(2005)\citenamefont {{Wang}},
  \citenamefont {{Haiman}}, \citenamefont {{Hu}}, \citenamefont {{Khoury}},\
  and\ \citenamefont {{May}}}]{wang05}%
  \BibitemOpen
  \bibfield  {author} {\bibinfo {author} {\bibfnamefont {S.}~\bibnamefont
  {{Wang}}}, \bibinfo {author} {\bibfnamefont {Z.}~\bibnamefont {{Haiman}}},
  \bibinfo {author} {\bibfnamefont {W.}~\bibnamefont {{Hu}}}, \bibinfo {author}
  {\bibfnamefont {J.}~\bibnamefont {{Khoury}}}, \ and\ \bibinfo {author}
  {\bibfnamefont {M.}~\bibnamefont {{May}}},\ }\href {\doibase
  10.1103/PhysRevLett.95.011302} {\bibfield  {journal} {\bibinfo  {journal}
  {\prl}\ }\textbf {\bibinfo {volume} {95}},\ \bibinfo {eid} {011302} (\bibinfo
  {year} {2005})},\ \Eprint {http://arxiv.org/abs/astro-ph/0505390}
  {arXiv:astro-ph/0505390 [astro-ph]} \BibitemShut {NoStop}%
\bibitem [{\citenamefont {{Mantz}}\ \emph {et~al.}(2008)\citenamefont
  {{Mantz}}, \citenamefont {{Allen}}, \citenamefont {{Ebeling}},\ and\
  \citenamefont {{Rapetti}}}]{mantz08}%
  \BibitemOpen
  \bibfield  {author} {\bibinfo {author} {\bibfnamefont {A.}~\bibnamefont
  {{Mantz}}}, \bibinfo {author} {\bibfnamefont {S.~W.}\ \bibnamefont
  {{Allen}}}, \bibinfo {author} {\bibfnamefont {H.}~\bibnamefont {{Ebeling}}},
  \ and\ \bibinfo {author} {\bibfnamefont {D.}~\bibnamefont {{Rapetti}}},\
  }\href {\doibase 10.1111/j.1365-2966.2008.13311.x} {\bibfield  {journal}
  {\bibinfo  {journal} {\mnras}\ }\textbf {\bibinfo {volume} {387}},\ \bibinfo
  {pages} {1179} (\bibinfo {year} {2008})},\ \Eprint
  {http://arxiv.org/abs/0709.4294} {arXiv:0709.4294} \BibitemShut {NoStop}%
\bibitem [{\citenamefont {{Vikhlinin}}\ \emph {et~al.}(2009)\citenamefont
  {{Vikhlinin}}, \citenamefont {{Kravtsov}}, \citenamefont {{Burenin}},
  \citenamefont {{Ebeling}}, \citenamefont {{Forman}}, \citenamefont
  {{Hornstrup}}, \citenamefont {{Jones}}, \citenamefont {{Murray}},
  \citenamefont {{Nagai}}, \citenamefont {{Quintana}},\ and\ \citenamefont
  {{Voevodkin}}}]{vikhlinin09}%
  \BibitemOpen
  \bibfield  {author} {\bibinfo {author} {\bibfnamefont {A.}~\bibnamefont
  {{Vikhlinin}}}, \bibinfo {author} {\bibfnamefont {A.~V.}\ \bibnamefont
  {{Kravtsov}}}, \bibinfo {author} {\bibfnamefont {R.~A.}\ \bibnamefont
  {{Burenin}}}, \bibinfo {author} {\bibfnamefont {H.}~\bibnamefont
  {{Ebeling}}}, \bibinfo {author} {\bibfnamefont {W.~R.}\ \bibnamefont
  {{Forman}}}, \bibinfo {author} {\bibfnamefont {A.}~\bibnamefont
  {{Hornstrup}}}, \bibinfo {author} {\bibfnamefont {C.}~\bibnamefont
  {{Jones}}}, \bibinfo {author} {\bibfnamefont {S.~S.}\ \bibnamefont
  {{Murray}}}, \bibinfo {author} {\bibfnamefont {D.}~\bibnamefont {{Nagai}}},
  \bibinfo {author} {\bibfnamefont {H.}~\bibnamefont {{Quintana}}}, \ and\
  \bibinfo {author} {\bibfnamefont {A.}~\bibnamefont {{Voevodkin}}},\ }\href
  {\doibase 10.1088/0004-637X/692/2/1060} {\bibfield  {journal} {\bibinfo
  {journal} {\apj}\ }\textbf {\bibinfo {volume} {692}},\ \bibinfo {pages}
  {1060} (\bibinfo {year} {2009})},\ \Eprint {http://arxiv.org/abs/0812.2720}
  {arXiv:0812.2720} \BibitemShut {NoStop}%
\bibitem [{\citenamefont {{Mantz}}\ \emph {et~al.}(2010)\citenamefont
  {{Mantz}}, \citenamefont {{Allen}},\ and\ \citenamefont
  {{Rapetti}}}]{mantz10c}%
  \BibitemOpen
  \bibfield  {author} {\bibinfo {author} {\bibfnamefont {A.}~\bibnamefont
  {{Mantz}}}, \bibinfo {author} {\bibfnamefont {S.~W.}\ \bibnamefont
  {{Allen}}}, \ and\ \bibinfo {author} {\bibfnamefont {D.}~\bibnamefont
  {{Rapetti}}},\ }\href {\doibase 10.1111/j.1365-2966.2010.16794.x} {\bibfield
  {journal} {\bibinfo  {journal} {\mnras}\ }\textbf {\bibinfo {volume} {406}},\
  \bibinfo {pages} {1805} (\bibinfo {year} {2010})},\ \Eprint
  {http://arxiv.org/abs/0911.1788} {arXiv:0911.1788 [astro-ph.CO]} \BibitemShut
  {NoStop}%
\bibitem [{\citenamefont {{Rozo}}\ \emph {et~al.}(2010)\citenamefont {{Rozo}},
  \citenamefont {{Wechsler}}, \citenamefont {{Rykoff}}, \citenamefont
  {{Annis}}, \citenamefont {{Becker}}, \citenamefont {{Evrard}}, \citenamefont
  {{Frieman}}, \citenamefont {{Hansen}}, \citenamefont {{Hao}}, \citenamefont
  {{Johnston}} \emph {et~al.}}]{rozo10}%
  \BibitemOpen
  \bibfield  {author} {\bibinfo {author} {\bibfnamefont {E.}~\bibnamefont
  {{Rozo}}}, \bibinfo {author} {\bibfnamefont {R.~H.}\ \bibnamefont
  {{Wechsler}}}, \bibinfo {author} {\bibfnamefont {E.~S.}\ \bibnamefont
  {{Rykoff}}}, \bibinfo {author} {\bibfnamefont {J.~T.}\ \bibnamefont
  {{Annis}}}, \bibinfo {author} {\bibfnamefont {M.~R.}\ \bibnamefont
  {{Becker}}}, \bibinfo {author} {\bibfnamefont {A.~E.}\ \bibnamefont
  {{Evrard}}}, \bibinfo {author} {\bibfnamefont {J.~A.}\ \bibnamefont
  {{Frieman}}}, \bibinfo {author} {\bibfnamefont {S.~M.}\ \bibnamefont
  {{Hansen}}}, \bibinfo {author} {\bibfnamefont {J.}~\bibnamefont {{Hao}}},
  \bibinfo {author} {\bibfnamefont {D.~E.}\ \bibnamefont {{Johnston}}},  \emph
  {et~al.},\ }\href {\doibase 10.1088/0004-637X/708/1/645} {\bibfield
  {journal} {\bibinfo  {journal} {\apj}\ }\textbf {\bibinfo {volume} {708}},\
  \bibinfo {pages} {645} (\bibinfo {year} {2010})},\ \Eprint
  {http://arxiv.org/abs/0902.3702} {arXiv:0902.3702 [astro-ph.CO]} \BibitemShut
  {NoStop}%
\bibitem [{\citenamefont {{Hasselfield}}\ \emph {et~al.}(2013)\citenamefont
  {{Hasselfield}}, \citenamefont {{Hilton}}, \citenamefont {{Marriage}},
  \citenamefont {{Addison}}, \citenamefont {{Barrientos}}, \citenamefont
  {{Battaglia}}, \citenamefont {{Battistelli}}, \citenamefont {{Bond}},
  \citenamefont {{Crichton}}, \citenamefont {{Das}} \emph
  {et~al.}}]{hasselfield13}%
  \BibitemOpen
  \bibfield  {author} {\bibinfo {author} {\bibfnamefont {M.}~\bibnamefont
  {{Hasselfield}}}, \bibinfo {author} {\bibfnamefont {M.}~\bibnamefont
  {{Hilton}}}, \bibinfo {author} {\bibfnamefont {T.~A.}\ \bibnamefont
  {{Marriage}}}, \bibinfo {author} {\bibfnamefont {G.~E.}\ \bibnamefont
  {{Addison}}}, \bibinfo {author} {\bibfnamefont {L.~F.}\ \bibnamefont
  {{Barrientos}}}, \bibinfo {author} {\bibfnamefont {N.}~\bibnamefont
  {{Battaglia}}}, \bibinfo {author} {\bibfnamefont {E.~S.}\ \bibnamefont
  {{Battistelli}}}, \bibinfo {author} {\bibfnamefont {J.~R.}\ \bibnamefont
  {{Bond}}}, \bibinfo {author} {\bibfnamefont {D.}~\bibnamefont {{Crichton}}},
  \bibinfo {author} {\bibfnamefont {S.}~\bibnamefont {{Das}}},  \emph
  {et~al.},\ }\href {\doibase 10.1088/1475-7516/2013/07/008} {\bibfield
  {journal} {\bibinfo  {journal} {\jcap}\ }\textbf {\bibinfo {volume} {7}},\
  \bibinfo {eid} {008} (\bibinfo {year} {2013})},\ \Eprint
  {http://arxiv.org/abs/1301.0816} {arXiv:1301.0816 [astro-ph.CO]} \BibitemShut
  {NoStop}%
\bibitem [{\citenamefont {{Mantz}}\ \emph {et~al.}(2015)\citenamefont
  {{Mantz}}, \citenamefont {{von der Linden}}, \citenamefont {{Allen}},
  \citenamefont {{Applegate}}, \citenamefont {{Kelly}}, \citenamefont
  {{Morris}}, \citenamefont {{Rapetti}}, \citenamefont {{Schmidt}},
  \citenamefont {{Adhikari}}, \citenamefont {{Allen}} \emph
  {et~al.}}]{mantz15}%
  \BibitemOpen
  \bibfield  {author} {\bibinfo {author} {\bibfnamefont {A.~B.}\ \bibnamefont
  {{Mantz}}}, \bibinfo {author} {\bibfnamefont {A.}~\bibnamefont {{von der
  Linden}}}, \bibinfo {author} {\bibfnamefont {S.~W.}\ \bibnamefont {{Allen}}},
  \bibinfo {author} {\bibfnamefont {D.~E.}\ \bibnamefont {{Applegate}}},
  \bibinfo {author} {\bibfnamefont {P.~L.}\ \bibnamefont {{Kelly}}}, \bibinfo
  {author} {\bibfnamefont {R.~G.}\ \bibnamefont {{Morris}}}, \bibinfo {author}
  {\bibfnamefont {D.~A.}\ \bibnamefont {{Rapetti}}}, \bibinfo {author}
  {\bibfnamefont {R.~W.}\ \bibnamefont {{Schmidt}}}, \bibinfo {author}
  {\bibfnamefont {S.}~\bibnamefont {{Adhikari}}}, \bibinfo {author}
  {\bibfnamefont {M.~T.}\ \bibnamefont {{Allen}}},  \emph {et~al.},\ }\href
  {\doibase 10.1093/mnras/stu2096} {\bibfield  {journal} {\bibinfo  {journal}
  {\mnras}\ }\textbf {\bibinfo {volume} {446}},\ \bibinfo {pages} {2205}
  (\bibinfo {year} {2015})},\ \Eprint {http://arxiv.org/abs/1407.4516}
  {arXiv:1407.4516} \BibitemShut {NoStop}%
\bibitem [{\citenamefont {{Planck Collaboration}}\ \emph
  {et~al.}(2016{\natexlab{a}})\citenamefont {{Planck Collaboration}},
  \citenamefont {{Ade}}, \citenamefont {{Aghanim}}, \citenamefont {{Arnaud}},
  \citenamefont {{Ashdown}}, \citenamefont {{Aumont}}, \citenamefont
  {{Baccigalupi}}, \citenamefont {{Banday}}, \citenamefont {{Barreiro}},
  \citenamefont {{Bartlett}} \emph {et~al.}}]{placksz15}%
  \BibitemOpen
  \bibfield  {author} {\bibinfo {author} {\bibnamefont {{Planck
  Collaboration}}}, \bibinfo {author} {\bibfnamefont {P.~A.~R.}\ \bibnamefont
  {{Ade}}}, \bibinfo {author} {\bibfnamefont {N.}~\bibnamefont {{Aghanim}}},
  \bibinfo {author} {\bibfnamefont {M.}~\bibnamefont {{Arnaud}}}, \bibinfo
  {author} {\bibfnamefont {M.}~\bibnamefont {{Ashdown}}}, \bibinfo {author}
  {\bibfnamefont {J.}~\bibnamefont {{Aumont}}}, \bibinfo {author}
  {\bibfnamefont {C.}~\bibnamefont {{Baccigalupi}}}, \bibinfo {author}
  {\bibfnamefont {A.~J.}\ \bibnamefont {{Banday}}}, \bibinfo {author}
  {\bibfnamefont {R.~B.}\ \bibnamefont {{Barreiro}}}, \bibinfo {author}
  {\bibfnamefont {J.~G.}\ \bibnamefont {{Bartlett}}},  \emph {et~al.},\ }\href
  {\doibase 10.1051/0004-6361/201525833} {\bibfield  {journal} {\bibinfo
  {journal} {\aap}\ }\textbf {\bibinfo {volume} {594}},\ \bibinfo {eid} {A24}
  (\bibinfo {year} {2016}{\natexlab{a}})},\ \Eprint
  {http://arxiv.org/abs/1502.01597} {arXiv:1502.01597} \BibitemShut {NoStop}%
\bibitem [{\citenamefont {{de Haan}}\ \emph {et~al.}(2016)\citenamefont {{de
  Haan}}, \citenamefont {{Benson}}, \citenamefont {{Bleem}}, \citenamefont
  {{Allen}}, \citenamefont {{Applegate}}, \citenamefont {{Ashby}},
  \citenamefont {{Bautz}}, \citenamefont {{Bayliss}}, \citenamefont
  {{Bocquet}}, \citenamefont {{Brodwin}} \emph {et~al.}}]{dehaan16}%
  \BibitemOpen
  \bibfield  {author} {\bibinfo {author} {\bibfnamefont {T.}~\bibnamefont {{de
  Haan}}}, \bibinfo {author} {\bibfnamefont {B.~A.}\ \bibnamefont {{Benson}}},
  \bibinfo {author} {\bibfnamefont {L.~E.}\ \bibnamefont {{Bleem}}}, \bibinfo
  {author} {\bibfnamefont {S.~W.}\ \bibnamefont {{Allen}}}, \bibinfo {author}
  {\bibfnamefont {D.~E.}\ \bibnamefont {{Applegate}}}, \bibinfo {author}
  {\bibfnamefont {M.~L.~N.}\ \bibnamefont {{Ashby}}}, \bibinfo {author}
  {\bibfnamefont {M.}~\bibnamefont {{Bautz}}}, \bibinfo {author} {\bibfnamefont
  {M.}~\bibnamefont {{Bayliss}}}, \bibinfo {author} {\bibfnamefont
  {S.}~\bibnamefont {{Bocquet}}}, \bibinfo {author} {\bibfnamefont
  {M.}~\bibnamefont {{Brodwin}}},  \emph {et~al.},\ }\href {\doibase
  10.3847/0004-637X/832/1/95} {\bibfield  {journal} {\bibinfo  {journal}
  {\apj}\ }\textbf {\bibinfo {volume} {832}},\ \bibinfo {eid} {95} (\bibinfo
  {year} {2016})},\ \Eprint {http://arxiv.org/abs/1603.06522}
  {arXiv:1603.06522} \BibitemShut {NoStop}%
\bibitem [{\citenamefont {{Salvati}}\ \emph {et~al.}(2018)\citenamefont
  {{Salvati}}, \citenamefont {{Douspis}},\ and\ \citenamefont
  {{Aghanim}}}]{salvati17}%
  \BibitemOpen
  \bibfield  {author} {\bibinfo {author} {\bibfnamefont {L.}~\bibnamefont
  {{Salvati}}}, \bibinfo {author} {\bibfnamefont {M.}~\bibnamefont
  {{Douspis}}}, \ and\ \bibinfo {author} {\bibfnamefont {N.}~\bibnamefont
  {{Aghanim}}},\ }\href {\doibase 10.1051/0004-6361/201731990} {\bibfield
  {journal} {\bibinfo  {journal} {\aap}\ }\textbf {\bibinfo {volume} {614}},\
  \bibinfo {eid} {A13} (\bibinfo {year} {2018})},\ \Eprint
  {http://arxiv.org/abs/1708.00697} {arXiv:1708.00697} \BibitemShut {NoStop}%
\bibitem [{\citenamefont {{Bocquet}}\ \emph {et~al.}(2019)\citenamefont
  {{Bocquet}}, \citenamefont {{Dietrich}}, \citenamefont {{Schrabback}},
  \citenamefont {{Bleem}}, \citenamefont {{Klein}}, \citenamefont {{Allen}},
  \citenamefont {{Applegate}}, \citenamefont {{Ashby}}, \citenamefont
  {{Bautz}},\ and\ \citenamefont {{Bayliss}}}]{bocquet19}%
  \BibitemOpen
  \bibfield  {author} {\bibinfo {author} {\bibfnamefont {S.}~\bibnamefont
  {{Bocquet}}}, \bibinfo {author} {\bibfnamefont {J.~P.}\ \bibnamefont
  {{Dietrich}}}, \bibinfo {author} {\bibfnamefont {T.}~\bibnamefont
  {{Schrabback}}}, \bibinfo {author} {\bibfnamefont {L.~E.}\ \bibnamefont
  {{Bleem}}}, \bibinfo {author} {\bibfnamefont {M.}~\bibnamefont {{Klein}}},
  \bibinfo {author} {\bibfnamefont {S.~W.}\ \bibnamefont {{Allen}}}, \bibinfo
  {author} {\bibfnamefont {D.~E.}\ \bibnamefont {{Applegate}}}, \bibinfo
  {author} {\bibfnamefont {M.~L.~N.}\ \bibnamefont {{Ashby}}}, \bibinfo
  {author} {\bibfnamefont {M.}~\bibnamefont {{Bautz}}}, \ and\ \bibinfo
  {author} {\bibfnamefont {M.}~\bibnamefont {{Bayliss}}},\ }\href {\doibase
  10.3847/1538-4357/ab1f10} {\bibfield  {journal} {\bibinfo  {journal} {\apj}\
  }\textbf {\bibinfo {volume} {878}},\ \bibinfo {eid} {55} (\bibinfo {year}
  {2019})},\ \Eprint {http://arxiv.org/abs/1812.01679} {arXiv:1812.01679
  [astro-ph.CO]} \BibitemShut {NoStop}%
\bibitem [{\citenamefont {{von der Linden}}\ \emph {et~al.}(2014)\citenamefont
  {{von der Linden}}, \citenamefont {{Allen}}, \citenamefont {{Applegate}},
  \citenamefont {{Kelly}}, \citenamefont {{Allen}}, \citenamefont {{Ebeling}},
  \citenamefont {{Burchat}}, \citenamefont {{Burke}}, \citenamefont
  {{Donovan}}, \citenamefont {{Morris}} \emph {et~al.}}]{linden14}%
  \BibitemOpen
  \bibfield  {author} {\bibinfo {author} {\bibfnamefont {A.}~\bibnamefont {{von
  der Linden}}}, \bibinfo {author} {\bibfnamefont {M.~T.}\ \bibnamefont
  {{Allen}}}, \bibinfo {author} {\bibfnamefont {D.~E.}\ \bibnamefont
  {{Applegate}}}, \bibinfo {author} {\bibfnamefont {P.~L.}\ \bibnamefont
  {{Kelly}}}, \bibinfo {author} {\bibfnamefont {S.~W.}\ \bibnamefont
  {{Allen}}}, \bibinfo {author} {\bibfnamefont {H.}~\bibnamefont {{Ebeling}}},
  \bibinfo {author} {\bibfnamefont {P.~R.}\ \bibnamefont {{Burchat}}}, \bibinfo
  {author} {\bibfnamefont {D.~L.}\ \bibnamefont {{Burke}}}, \bibinfo {author}
  {\bibfnamefont {D.}~\bibnamefont {{Donovan}}}, \bibinfo {author}
  {\bibfnamefont {R.~G.}\ \bibnamefont {{Morris}}},  \emph {et~al.},\ }\href
  {\doibase 10.1093/mnras/stt1945} {\bibfield  {journal} {\bibinfo  {journal}
  {\mnras}\ }\textbf {\bibinfo {volume} {439}},\ \bibinfo {pages} {2} (\bibinfo
  {year} {2014})},\ \Eprint {http://arxiv.org/abs/1208.0597} {arXiv:1208.0597}
  \BibitemShut {NoStop}%
\bibitem [{\citenamefont {{LSST Science Collaboration}}\ \emph
  {et~al.}(2009)\citenamefont {{LSST Science Collaboration}}, \citenamefont
  {{Abell}}, \citenamefont {{Allison}}, \citenamefont {{Anderson}},
  \citenamefont {{Andrew}}, \citenamefont {{Angel}}, \citenamefont {{Armus}},
  \citenamefont {{Arnett}}, \citenamefont {{Asztalos}}, \citenamefont
  {{Axelrod}},\ and\ \citenamefont {et~al.}}]{lsst09}%
  \BibitemOpen
  \bibfield  {author} {\bibinfo {author} {\bibnamefont {{LSST Science
  Collaboration}}}, \bibinfo {author} {\bibfnamefont {P.~A.}\ \bibnamefont
  {{Abell}}}, \bibinfo {author} {\bibfnamefont {J.}~\bibnamefont {{Allison}}},
  \bibinfo {author} {\bibfnamefont {S.~F.}\ \bibnamefont {{Anderson}}},
  \bibinfo {author} {\bibfnamefont {J.~R.}\ \bibnamefont {{Andrew}}}, \bibinfo
  {author} {\bibfnamefont {J.~R.~P.}\ \bibnamefont {{Angel}}}, \bibinfo
  {author} {\bibfnamefont {L.}~\bibnamefont {{Armus}}}, \bibinfo {author}
  {\bibfnamefont {D.}~\bibnamefont {{Arnett}}}, \bibinfo {author}
  {\bibfnamefont {S.~J.}\ \bibnamefont {{Asztalos}}}, \bibinfo {author}
  {\bibfnamefont {T.~S.}\ \bibnamefont {{Axelrod}}}, \ and\ \bibinfo {author}
  {\bibnamefont {et~al.}},\ }\href@noop {} {\bibfield  {journal} {\bibinfo
  {journal} {ArXiv e-prints}\ } (\bibinfo {year} {2009})},\ \Eprint
  {http://arxiv.org/abs/0912.0201} {arXiv:0912.0201 [astro-ph.IM]} \BibitemShut
  {NoStop}%
\bibitem [{\citenamefont {{Merloni}}\ \emph {et~al.}(2012)\citenamefont
  {{Merloni}}, \citenamefont {{Predehl}}, \citenamefont {{Becker}},
  \citenamefont {{B{\"o}hringer}}, \citenamefont {{Boller}}, \citenamefont
  {{Brunner}}, \citenamefont {{Brusa}}, \citenamefont {{Dennerl}},
  \citenamefont {{Freyberg}}, \citenamefont {{Friedrich}} \emph
  {et~al.}}]{erosita12}%
  \BibitemOpen
  \bibfield  {author} {\bibinfo {author} {\bibfnamefont {A.}~\bibnamefont
  {{Merloni}}}, \bibinfo {author} {\bibfnamefont {P.}~\bibnamefont
  {{Predehl}}}, \bibinfo {author} {\bibfnamefont {W.}~\bibnamefont {{Becker}}},
  \bibinfo {author} {\bibfnamefont {H.}~\bibnamefont {{B{\"o}hringer}}},
  \bibinfo {author} {\bibfnamefont {T.}~\bibnamefont {{Boller}}}, \bibinfo
  {author} {\bibfnamefont {H.}~\bibnamefont {{Brunner}}}, \bibinfo {author}
  {\bibfnamefont {M.}~\bibnamefont {{Brusa}}}, \bibinfo {author} {\bibfnamefont
  {K.}~\bibnamefont {{Dennerl}}}, \bibinfo {author} {\bibfnamefont
  {M.}~\bibnamefont {{Freyberg}}}, \bibinfo {author} {\bibfnamefont
  {P.}~\bibnamefont {{Friedrich}}},  \emph {et~al.},\ }\href@noop {} {\bibfield
   {journal} {\bibinfo  {journal} {ArXiv e-prints}\ } (\bibinfo {year}
  {2012})},\ \Eprint {http://arxiv.org/abs/1209.3114} {arXiv:1209.3114
  [astro-ph.HE]} \BibitemShut {NoStop}%
\bibitem [{\citenamefont {{CMB-S4 Collaboration}}\ \emph
  {et~al.}(2016)\citenamefont {{CMB-S4 Collaboration}}, \citenamefont
  {{Abazajian}}, \citenamefont {{Adshead}}, \citenamefont {{Ahmed}},
  \citenamefont {{Allen}}, \citenamefont {{Alonso}}, \citenamefont {{Arnold}},
  \citenamefont {{Baccigalupi}}, \citenamefont {{Bartlett}}, \citenamefont
  {{Battaglia}} \emph {et~al.}}]{cmbs4-sb1}%
  \BibitemOpen
  \bibfield  {author} {\bibinfo {author} {\bibnamefont {{CMB-S4
  Collaboration}}}, \bibinfo {author} {\bibfnamefont {K.~N.}\ \bibnamefont
  {{Abazajian}}}, \bibinfo {author} {\bibfnamefont {P.}~\bibnamefont
  {{Adshead}}}, \bibinfo {author} {\bibfnamefont {Z.}~\bibnamefont {{Ahmed}}},
  \bibinfo {author} {\bibfnamefont {S.~W.}\ \bibnamefont {{Allen}}}, \bibinfo
  {author} {\bibfnamefont {D.}~\bibnamefont {{Alonso}}}, \bibinfo {author}
  {\bibfnamefont {K.~S.}\ \bibnamefont {{Arnold}}}, \bibinfo {author}
  {\bibfnamefont {C.}~\bibnamefont {{Baccigalupi}}}, \bibinfo {author}
  {\bibfnamefont {J.~G.}\ \bibnamefont {{Bartlett}}}, \bibinfo {author}
  {\bibfnamefont {N.}~\bibnamefont {{Battaglia}}},  \emph {et~al.},\
  }\href@noop {} {\bibfield  {journal} {\bibinfo  {journal} {ArXiv e-prints}\ }
  (\bibinfo {year} {2016})},\ \Eprint {http://arxiv.org/abs/1610.02743}
  {arXiv:1610.02743} \BibitemShut {NoStop}%
\bibitem [{\citenamefont {{Lewis}}\ and\ \citenamefont
  {{King}}(2006)}]{lewis06a}%
  \BibitemOpen
  \bibfield  {author} {\bibinfo {author} {\bibfnamefont {A.}~\bibnamefont
  {{Lewis}}}\ and\ \bibinfo {author} {\bibfnamefont {L.}~\bibnamefont
  {{King}}},\ }\href {\doibase 10.1103/PhysRevD.73.063006} {\bibfield
  {journal} {\bibinfo  {journal} {\prd}\ }\textbf {\bibinfo {volume} {73}},\
  \bibinfo {eid} {063006} (\bibinfo {year} {2006})},\ \Eprint
  {http://arxiv.org/abs/astro-ph/0512104} {arXiv:astro-ph/0512104 [astro-ph]}
  \BibitemShut {NoStop}%
\bibitem [{\citenamefont {{Madhavacheril}}\ \emph {et~al.}(2015)\citenamefont
  {{Madhavacheril}}, \citenamefont {{Sehgal}}, \citenamefont {{Allison}},
  \citenamefont {{Battaglia}}, \citenamefont {{Bond}}, \citenamefont
  {{Calabrese}}, \citenamefont {{Caliguiri}}, \citenamefont {{Coughlin}},
  \citenamefont {{Crichton}}, \citenamefont {{Datta}} \emph
  {et~al.}}]{madhavacheril15}%
  \BibitemOpen
  \bibfield  {author} {\bibinfo {author} {\bibfnamefont {M.}~\bibnamefont
  {{Madhavacheril}}}, \bibinfo {author} {\bibfnamefont {N.}~\bibnamefont
  {{Sehgal}}}, \bibinfo {author} {\bibfnamefont {R.}~\bibnamefont {{Allison}}},
  \bibinfo {author} {\bibfnamefont {N.}~\bibnamefont {{Battaglia}}}, \bibinfo
  {author} {\bibfnamefont {J.~R.}\ \bibnamefont {{Bond}}}, \bibinfo {author}
  {\bibfnamefont {E.}~\bibnamefont {{Calabrese}}}, \bibinfo {author}
  {\bibfnamefont {J.}~\bibnamefont {{Caliguiri}}}, \bibinfo {author}
  {\bibfnamefont {K.}~\bibnamefont {{Coughlin}}}, \bibinfo {author}
  {\bibfnamefont {D.}~\bibnamefont {{Crichton}}}, \bibinfo {author}
  {\bibfnamefont {R.}~\bibnamefont {{Datta}}},  \emph {et~al.},\ }\href
  {\doibase 10.1103/PhysRevLett.114.151302} {\bibfield  {journal} {\bibinfo
  {journal} {Physical Review Letters}\ }\textbf {\bibinfo {volume} {114}},\
  \bibinfo {eid} {151302} (\bibinfo {year} {2015})},\ \Eprint
  {http://arxiv.org/abs/1411.7999} {arXiv:1411.7999} \BibitemShut {NoStop}%
\bibitem [{\citenamefont {{Baxter}}\ \emph {et~al.}(2015)\citenamefont
  {{Baxter}}, \citenamefont {{Keisler}}, \citenamefont {{Dodelson}},
  \citenamefont {{Aird}}, \citenamefont {{Allen}}, \citenamefont {{Ashby}},
  \citenamefont {{Bautz}}, \citenamefont {{Bayliss}}, \citenamefont {{Benson}},
  \citenamefont {{Bleem}} \emph {et~al.}}]{baxter15}%
  \BibitemOpen
  \bibfield  {author} {\bibinfo {author} {\bibfnamefont {E.~J.}\ \bibnamefont
  {{Baxter}}}, \bibinfo {author} {\bibfnamefont {R.}~\bibnamefont {{Keisler}}},
  \bibinfo {author} {\bibfnamefont {S.}~\bibnamefont {{Dodelson}}}, \bibinfo
  {author} {\bibfnamefont {K.~A.}\ \bibnamefont {{Aird}}}, \bibinfo {author}
  {\bibfnamefont {S.~W.}\ \bibnamefont {{Allen}}}, \bibinfo {author}
  {\bibfnamefont {M.~L.~N.}\ \bibnamefont {{Ashby}}}, \bibinfo {author}
  {\bibfnamefont {M.}~\bibnamefont {{Bautz}}}, \bibinfo {author} {\bibfnamefont
  {M.}~\bibnamefont {{Bayliss}}}, \bibinfo {author} {\bibfnamefont {B.~A.}\
  \bibnamefont {{Benson}}}, \bibinfo {author} {\bibfnamefont {L.~E.}\
  \bibnamefont {{Bleem}}},  \emph {et~al.},\ }\href {\doibase
  10.1088/0004-637X/806/2/247} {\bibfield  {journal} {\bibinfo  {journal}
  {\apj}\ }\textbf {\bibinfo {volume} {806}},\ \bibinfo {eid} {247} (\bibinfo
  {year} {2015})},\ \Eprint {http://arxiv.org/abs/1412.7521} {arXiv:1412.7521}
  \BibitemShut {NoStop}%
\bibitem [{\citenamefont {{Geach}}\ and\ \citenamefont
  {{Peacock}}(2017)}]{geach17}%
  \BibitemOpen
  \bibfield  {author} {\bibinfo {author} {\bibfnamefont {J.~E.}\ \bibnamefont
  {{Geach}}}\ and\ \bibinfo {author} {\bibfnamefont {J.~A.}\ \bibnamefont
  {{Peacock}}},\ }\href {\doibase 10.1038/s41550-017-0259-1} {\bibfield
  {journal} {\bibinfo  {journal} {Nature Astronomy}\ }\textbf {\bibinfo
  {volume} {1}},\ \bibinfo {pages} {795} (\bibinfo {year} {2017})},\ \Eprint
  {http://arxiv.org/abs/1707.09369} {arXiv:1707.09369} \BibitemShut {NoStop}%
\bibitem [{\citenamefont {{Baxter}}\ \emph {et~al.}(2018)\citenamefont
  {{Baxter}}, \citenamefont {{Raghunathan}}, \citenamefont {{Crawford}},
  \citenamefont {{Fosalba}}, \citenamefont {{Hou}}, \citenamefont {{Holder}},
  \citenamefont {{Omori}}, \citenamefont {{Patil}}, \citenamefont {{Rozo}},
  \citenamefont {{Abbott}} \emph {et~al.}}]{baxter18}%
  \BibitemOpen
  \bibfield  {author} {\bibinfo {author} {\bibfnamefont {E.~J.}\ \bibnamefont
  {{Baxter}}}, \bibinfo {author} {\bibfnamefont {S.}~\bibnamefont
  {{Raghunathan}}}, \bibinfo {author} {\bibfnamefont {T.~M.}\ \bibnamefont
  {{Crawford}}}, \bibinfo {author} {\bibfnamefont {P.}~\bibnamefont
  {{Fosalba}}}, \bibinfo {author} {\bibfnamefont {Z.}~\bibnamefont {{Hou}}},
  \bibinfo {author} {\bibfnamefont {G.~P.}\ \bibnamefont {{Holder}}}, \bibinfo
  {author} {\bibfnamefont {Y.}~\bibnamefont {{Omori}}}, \bibinfo {author}
  {\bibfnamefont {S.}~\bibnamefont {{Patil}}}, \bibinfo {author} {\bibfnamefont
  {E.}~\bibnamefont {{Rozo}}}, \bibinfo {author} {\bibfnamefont {T.~M.~C.}\
  \bibnamefont {{Abbott}}},  \emph {et~al.},\ }\href {\doibase
  10.1093/mnras/sty305} {\bibfield  {journal} {\bibinfo  {journal} {\mnras}\ }
  (\bibinfo {year} {2018}),\ 10.1093/mnras/sty305},\ \Eprint
  {http://arxiv.org/abs/1708.01360} {arXiv:1708.01360} \BibitemShut {NoStop}%
\bibitem [{\citenamefont {{Raghunathan}}\ \emph {et~al.}(2018)\citenamefont
  {{Raghunathan}}, \citenamefont {{Bianchini}},\ and\ \citenamefont
  {{Reichardt}}}]{raghunathan17b}%
  \BibitemOpen
  \bibfield  {author} {\bibinfo {author} {\bibfnamefont {S.}~\bibnamefont
  {{Raghunathan}}}, \bibinfo {author} {\bibfnamefont {F.}~\bibnamefont
  {{Bianchini}}}, \ and\ \bibinfo {author} {\bibfnamefont {C.~L.}\ \bibnamefont
  {{Reichardt}}},\ }\href {\doibase 10.1103/PhysRevD.98.043506} {\bibfield
  {journal} {\bibinfo  {journal} {\prd}\ }\textbf {\bibinfo {volume} {98}},\
  \bibinfo {eid} {043506} (\bibinfo {year} {2018})},\ \Eprint
  {http://arxiv.org/abs/1710.09770} {arXiv:1710.09770} \BibitemShut {NoStop}%
\bibitem [{\citenamefont {{Raghunathan}}\ \emph {et~al.}(2019)\citenamefont
  {{Raghunathan}}, \citenamefont {{Patil}}, \citenamefont {{Baxter}},
  \citenamefont {{Benson}}, \citenamefont {{Bleem}}, \citenamefont {{Chou}},
  \citenamefont {{Crawford}}, \citenamefont {{Holder}}, \citenamefont
  {{McClintock}}, \citenamefont {{Reichardt}} \emph {et~al.}}]{raghunathan18}%
  \BibitemOpen
  \bibfield  {author} {\bibinfo {author} {\bibfnamefont {S.}~\bibnamefont
  {{Raghunathan}}}, \bibinfo {author} {\bibfnamefont {S.}~\bibnamefont
  {{Patil}}}, \bibinfo {author} {\bibfnamefont {E.}~\bibnamefont {{Baxter}}},
  \bibinfo {author} {\bibfnamefont {B.~A.}\ \bibnamefont {{Benson}}}, \bibinfo
  {author} {\bibfnamefont {L.~E.}\ \bibnamefont {{Bleem}}}, \bibinfo {author}
  {\bibfnamefont {T.~L.}\ \bibnamefont {{Chou}}}, \bibinfo {author}
  {\bibfnamefont {T.~M.}\ \bibnamefont {{Crawford}}}, \bibinfo {author}
  {\bibfnamefont {G.~P.}\ \bibnamefont {{Holder}}}, \bibinfo {author}
  {\bibfnamefont {T.}~\bibnamefont {{McClintock}}}, \bibinfo {author}
  {\bibfnamefont {C.~L.}\ \bibnamefont {{Reichardt}}},  \emph {et~al.},\ }\href
  {\doibase 10.3847/1538-4357/ab01ca} {\bibfield  {journal} {\bibinfo
  {journal} {\apj}\ }\textbf {\bibinfo {volume} {872}},\ \bibinfo {eid} {170}
  (\bibinfo {year} {2019})},\ \Eprint {http://arxiv.org/abs/1810.10998}
  {arXiv:1810.10998 [astro-ph.CO]} \BibitemShut {NoStop}%
\bibitem [{\citenamefont {{Raghunathan}}\ \emph {et~al.}(2017)\citenamefont
  {{Raghunathan}}, \citenamefont {{Patil}}, \citenamefont {{Baxter}},
  \citenamefont {{Bianchini}}, \citenamefont {{Bleem}}, \citenamefont
  {{Crawford}}, \citenamefont {{Holder}}, \citenamefont {{Manzotti}},\ and\
  \citenamefont {{Reichardt}}}]{raghunathan17a}%
  \BibitemOpen
  \bibfield  {author} {\bibinfo {author} {\bibfnamefont {S.}~\bibnamefont
  {{Raghunathan}}}, \bibinfo {author} {\bibfnamefont {S.}~\bibnamefont
  {{Patil}}}, \bibinfo {author} {\bibfnamefont {E.~J.}\ \bibnamefont
  {{Baxter}}}, \bibinfo {author} {\bibfnamefont {F.}~\bibnamefont
  {{Bianchini}}}, \bibinfo {author} {\bibfnamefont {L.~E.}\ \bibnamefont
  {{Bleem}}}, \bibinfo {author} {\bibfnamefont {T.~M.}\ \bibnamefont
  {{Crawford}}}, \bibinfo {author} {\bibfnamefont {G.~P.}\ \bibnamefont
  {{Holder}}}, \bibinfo {author} {\bibfnamefont {A.}~\bibnamefont
  {{Manzotti}}}, \ and\ \bibinfo {author} {\bibfnamefont {C.~L.}\ \bibnamefont
  {{Reichardt}}},\ }\href {\doibase 10.1088/1475-7516/2017/08/030} {\bibfield
  {journal} {\bibinfo  {journal} {\jcap}\ }\textbf {\bibinfo {volume} {8}},\
  \bibinfo {eid} {030} (\bibinfo {year} {2017})},\ \Eprint
  {http://arxiv.org/abs/1705.00411} {arXiv:1705.00411} \BibitemShut {NoStop}%
\bibitem [{\citenamefont {{Hu}}\ \emph {et~al.}(2007)\citenamefont {{Hu}},
  \citenamefont {{DeDeo}},\ and\ \citenamefont {{Vale}}}]{hu07}%
  \BibitemOpen
  \bibfield  {author} {\bibinfo {author} {\bibfnamefont {W.}~\bibnamefont
  {{Hu}}}, \bibinfo {author} {\bibfnamefont {S.}~\bibnamefont {{DeDeo}}}, \
  and\ \bibinfo {author} {\bibfnamefont {C.}~\bibnamefont {{Vale}}},\ }\href
  {\doibase 10.1088/1367-2630/9/12/441} {\bibfield  {journal} {\bibinfo
  {journal} {New Journal of Physics}\ }\textbf {\bibinfo {volume} {9}},\
  \bibinfo {pages} {441} (\bibinfo {year} {2007})},\ \Eprint
  {http://arxiv.org/abs/arXiv:astro-ph/0701276} {arXiv:astro-ph/0701276}
  \BibitemShut {NoStop}%
\bibitem [{\citenamefont {{Yoo}}\ \emph {et~al.}(2010)\citenamefont {{Yoo}},
  \citenamefont {{Zaldarriaga}},\ and\ \citenamefont {{Hernquist}}}]{yoo10}%
  \BibitemOpen
  \bibfield  {author} {\bibinfo {author} {\bibfnamefont {J.}~\bibnamefont
  {{Yoo}}}, \bibinfo {author} {\bibfnamefont {M.}~\bibnamefont
  {{Zaldarriaga}}}, \ and\ \bibinfo {author} {\bibfnamefont {L.}~\bibnamefont
  {{Hernquist}}},\ }\href {\doibase 10.1103/PhysRevD.81.123006} {\bibfield
  {journal} {\bibinfo  {journal} {\prd}\ }\textbf {\bibinfo {volume} {81}},\
  \bibinfo {eid} {123006} (\bibinfo {year} {2010})},\ \Eprint
  {http://arxiv.org/abs/1005.0847} {arXiv:1005.0847} \BibitemShut {NoStop}%
\bibitem [{\citenamefont {{Dodelson}}(2004)}]{dodelson04}%
  \BibitemOpen
  \bibfield  {author} {\bibinfo {author} {\bibfnamefont {S.}~\bibnamefont
  {{Dodelson}}},\ }\href {\doibase 10.1103/PhysRevD.70.023009} {\bibfield
  {journal} {\bibinfo  {journal} {\prd}\ }\textbf {\bibinfo {volume} {70}},\
  \bibinfo {eid} {023009} (\bibinfo {year} {2004})},\ \Eprint
  {http://arxiv.org/abs/arXiv:astro-ph/0402314} {arXiv:astro-ph/0402314}
  \BibitemShut {NoStop}%
\bibitem [{\citenamefont {{Planck Collaboration}}\ \emph
  {et~al.}(2016{\natexlab{b}})\citenamefont {{Planck Collaboration}},
  \citenamefont {{Ade}}, \citenamefont {{Aghanim}}, \citenamefont {{Arnaud}},
  \citenamefont {{Ashdown}}, \citenamefont {{Aumont}}, \citenamefont
  {{Baccigalupi}}, \citenamefont {{Banday}}, \citenamefont {{Barreiro}},
  \citenamefont {{Bartlett}},\ and\ \citenamefont {et~al.}}]{planck15-13}%
  \BibitemOpen
  \bibfield  {author} {\bibinfo {author} {\bibnamefont {{Planck
  Collaboration}}}, \bibinfo {author} {\bibfnamefont {P.~A.~R.}\ \bibnamefont
  {{Ade}}}, \bibinfo {author} {\bibfnamefont {N.}~\bibnamefont {{Aghanim}}},
  \bibinfo {author} {\bibfnamefont {M.}~\bibnamefont {{Arnaud}}}, \bibinfo
  {author} {\bibfnamefont {M.}~\bibnamefont {{Ashdown}}}, \bibinfo {author}
  {\bibfnamefont {J.}~\bibnamefont {{Aumont}}}, \bibinfo {author}
  {\bibfnamefont {C.}~\bibnamefont {{Baccigalupi}}}, \bibinfo {author}
  {\bibfnamefont {A.~J.}\ \bibnamefont {{Banday}}}, \bibinfo {author}
  {\bibfnamefont {R.~B.}\ \bibnamefont {{Barreiro}}}, \bibinfo {author}
  {\bibfnamefont {J.~G.}\ \bibnamefont {{Bartlett}}}, \ and\ \bibinfo {author}
  {\bibnamefont {et~al.}},\ }\href {\doibase 10.1051/0004-6361/201525830}
  {\bibfield  {journal} {\bibinfo  {journal} {\aap}\ }\textbf {\bibinfo
  {volume} {594}},\ \bibinfo {eid} {A13} (\bibinfo {year}
  {2016}{\natexlab{b}})},\ \Eprint {http://arxiv.org/abs/1502.01589}
  {arXiv:1502.01589} \BibitemShut {NoStop}%
\bibitem [{\citenamefont {{Lewis}}\ \emph {et~al.}(2000)\citenamefont
  {{Lewis}}, \citenamefont {{Challinor}},\ and\ \citenamefont
  {{Lasenby}}}]{lewis00}%
  \BibitemOpen
  \bibfield  {author} {\bibinfo {author} {\bibfnamefont {A.}~\bibnamefont
  {{Lewis}}}, \bibinfo {author} {\bibfnamefont {A.}~\bibnamefont
  {{Challinor}}}, \ and\ \bibinfo {author} {\bibfnamefont {A.}~\bibnamefont
  {{Lasenby}}},\ }\href {\doibase 10.1086/309179} {\bibfield  {journal}
  {\bibinfo  {journal} {\apj}\ }\textbf {\bibinfo {volume} {538}},\ \bibinfo
  {pages} {473} (\bibinfo {year} {2000})}\BibitemShut {NoStop}%
\bibitem [{\citenamefont {Padin}\ \emph {et~al.}(2008)\citenamefont {Padin},
  \citenamefont {Staniszewski}, \citenamefont {Keisler}, \citenamefont {Joy},
  \citenamefont {Stark}, \citenamefont {Ade}, \citenamefont {Aird},
  \citenamefont {Benson}, \citenamefont {Bleem}, \citenamefont {Carlstrom}
  \emph {et~al.}}]{padin08}%
  \BibitemOpen
  \bibfield  {author} {\bibinfo {author} {\bibfnamefont {S.}~\bibnamefont
  {Padin}}, \bibinfo {author} {\bibfnamefont {Z.}~\bibnamefont {Staniszewski}},
  \bibinfo {author} {\bibfnamefont {R.}~\bibnamefont {Keisler}}, \bibinfo
  {author} {\bibfnamefont {M.}~\bibnamefont {Joy}}, \bibinfo {author}
  {\bibfnamefont {A.~A.}\ \bibnamefont {Stark}}, \bibinfo {author}
  {\bibfnamefont {P.~A.~R.}\ \bibnamefont {Ade}}, \bibinfo {author}
  {\bibfnamefont {K.~A.}\ \bibnamefont {Aird}}, \bibinfo {author}
  {\bibfnamefont {B.~A.}\ \bibnamefont {Benson}}, \bibinfo {author}
  {\bibfnamefont {L.~E.}\ \bibnamefont {Bleem}}, \bibinfo {author}
  {\bibfnamefont {J.~E.}\ \bibnamefont {Carlstrom}},  \emph {et~al.},\
  }\href@noop {} {\bibfield  {journal} {\bibinfo  {journal} {Appl. Opt.}\
  }\textbf {\bibinfo {volume} {47}},\ \bibinfo {pages} {4418} (\bibinfo {year}
  {2008})}\BibitemShut {NoStop}%
\bibitem [{\citenamefont {{Carlstrom}}\ \emph {et~al.}(2011)\citenamefont
  {{Carlstrom}}, \citenamefont {{Ade}}, \citenamefont {{Aird}}, \citenamefont
  {{Benson}}, \citenamefont {{Bleem}}, \citenamefont {{Busetti}}, \citenamefont
  {{Chang}}, \citenamefont {{Chauvin}}, \citenamefont {{Cho}}, \citenamefont
  {{Crawford}} \emph {et~al.}}]{carlstrom11}%
  \BibitemOpen
  \bibfield  {author} {\bibinfo {author} {\bibfnamefont {J.~E.}\ \bibnamefont
  {{Carlstrom}}}, \bibinfo {author} {\bibfnamefont {P.~A.~R.}\ \bibnamefont
  {{Ade}}}, \bibinfo {author} {\bibfnamefont {K.~A.}\ \bibnamefont {{Aird}}},
  \bibinfo {author} {\bibfnamefont {B.~A.}\ \bibnamefont {{Benson}}}, \bibinfo
  {author} {\bibfnamefont {L.~E.}\ \bibnamefont {{Bleem}}}, \bibinfo {author}
  {\bibfnamefont {S.}~\bibnamefont {{Busetti}}}, \bibinfo {author}
  {\bibfnamefont {C.~L.}\ \bibnamefont {{Chang}}}, \bibinfo {author}
  {\bibfnamefont {E.}~\bibnamefont {{Chauvin}}}, \bibinfo {author}
  {\bibfnamefont {H.-M.}\ \bibnamefont {{Cho}}}, \bibinfo {author}
  {\bibfnamefont {T.~M.}\ \bibnamefont {{Crawford}}},  \emph {et~al.},\ }\href
  {\doibase 10.1086/659879} {\bibfield  {journal} {\bibinfo  {journal} {\pasp}\
  }\textbf {\bibinfo {volume} {123}},\ \bibinfo {pages} {568} (\bibinfo {year}
  {2011})},\ \Eprint {http://arxiv.org/abs/0907.4445} {arXiv:0907.4445}
  \BibitemShut {NoStop}%
\bibitem [{\citenamefont {{Austermann}}\ \emph {et~al.}(2012)\citenamefont
  {{Austermann}}, \citenamefont {{Aird}}, \citenamefont {{Beall}},
  \citenamefont {{Becker}}, \citenamefont {{Bender}}, \citenamefont {{Benson}},
  \citenamefont {{Bleem}}, \citenamefont {{Britton}}, \citenamefont
  {{Carlstrom}}, \citenamefont {{Chang}} \emph {et~al.}}]{austermann12}%
  \BibitemOpen
  \bibfield  {author} {\bibinfo {author} {\bibfnamefont {J.~E.}\ \bibnamefont
  {{Austermann}}}, \bibinfo {author} {\bibfnamefont {K.~A.}\ \bibnamefont
  {{Aird}}}, \bibinfo {author} {\bibfnamefont {J.~A.}\ \bibnamefont {{Beall}}},
  \bibinfo {author} {\bibfnamefont {D.}~\bibnamefont {{Becker}}}, \bibinfo
  {author} {\bibfnamefont {A.}~\bibnamefont {{Bender}}}, \bibinfo {author}
  {\bibfnamefont {B.~A.}\ \bibnamefont {{Benson}}}, \bibinfo {author}
  {\bibfnamefont {L.~E.}\ \bibnamefont {{Bleem}}}, \bibinfo {author}
  {\bibfnamefont {J.}~\bibnamefont {{Britton}}}, \bibinfo {author}
  {\bibfnamefont {J.~E.}\ \bibnamefont {{Carlstrom}}}, \bibinfo {author}
  {\bibfnamefont {C.~L.}\ \bibnamefont {{Chang}}},  \emph {et~al.},\ }in\ \href
  {\doibase 10.1117/12.927286} {\emph {\bibinfo {booktitle} {Society of
  Photo-Optical Instrumentation Engineers (SPIE) Conference Series}}},\ Vol.\
  \bibinfo {volume} {8452}\ (\bibinfo {year} {2012})\ \Eprint
  {http://arxiv.org/abs/1210.4970} {arXiv:1210.4970 [astro-ph.IM]} \BibitemShut
  {NoStop}%
\bibitem [{\citenamefont {{Sunyaev}}\ and\ \citenamefont
  {{Zel'dovich}}(1972)}]{sunyaev72}%
  \BibitemOpen
  \bibfield  {author} {\bibinfo {author} {\bibfnamefont {R.~A.}\ \bibnamefont
  {{Sunyaev}}}\ and\ \bibinfo {author} {\bibfnamefont {Y.~B.}\ \bibnamefont
  {{Zel'dovich}}},\ }\href@noop {} {\bibfield  {journal} {\bibinfo  {journal}
  {Comments on Astrophysics and Space Physics}\ }\textbf {\bibinfo {volume}
  {4}},\ \bibinfo {pages} {173} (\bibinfo {year} {1972})}\BibitemShut {NoStop}%
\bibitem [{\citenamefont {{Sunyaev}}\ and\ \citenamefont
  {{Zeldovich}}(1980)}]{sunyaev80b}%
  \BibitemOpen
  \bibfield  {author} {\bibinfo {author} {\bibfnamefont {R.~A.}\ \bibnamefont
  {{Sunyaev}}}\ and\ \bibinfo {author} {\bibfnamefont {Y.~B.}\ \bibnamefont
  {{Zeldovich}}},\ }\href
  {http://adsabs.harvard.edu/cgi-bin/nph-bib_query?bibcode=1980MNRAS.190..413S&db_key=AST}
  {\bibfield  {journal} {\bibinfo  {journal} {\mnras}\ }\textbf {\bibinfo
  {volume} {190}},\ \bibinfo {pages} {413} (\bibinfo {year}
  {1980})}\BibitemShut {NoStop}%
\bibitem [{\citenamefont {{Henning}}\ \emph {et~al.}(2018)\citenamefont
  {{Henning}}, \citenamefont {{Sayre}}, \citenamefont {{Reichardt}},
  \citenamefont {{Ade}}, \citenamefont {{Anderson}}, \citenamefont
  {{Austermann}}, \citenamefont {{Beall}}, \citenamefont {{Bender}},
  \citenamefont {{Benson}}, \citenamefont {{Bleem}} \emph
  {et~al.}}]{henning18}%
  \BibitemOpen
  \bibfield  {author} {\bibinfo {author} {\bibfnamefont {J.~W.}\ \bibnamefont
  {{Henning}}}, \bibinfo {author} {\bibfnamefont {J.~T.}\ \bibnamefont
  {{Sayre}}}, \bibinfo {author} {\bibfnamefont {C.~L.}\ \bibnamefont
  {{Reichardt}}}, \bibinfo {author} {\bibfnamefont {P.~A.~R.}\ \bibnamefont
  {{Ade}}}, \bibinfo {author} {\bibfnamefont {A.~J.}\ \bibnamefont
  {{Anderson}}}, \bibinfo {author} {\bibfnamefont {J.~E.}\ \bibnamefont
  {{Austermann}}}, \bibinfo {author} {\bibfnamefont {J.~A.}\ \bibnamefont
  {{Beall}}}, \bibinfo {author} {\bibfnamefont {A.~N.}\ \bibnamefont
  {{Bender}}}, \bibinfo {author} {\bibfnamefont {B.~A.}\ \bibnamefont
  {{Benson}}}, \bibinfo {author} {\bibfnamefont {L.~E.}\ \bibnamefont
  {{Bleem}}},  \emph {et~al.},\ }\href {\doibase 10.3847/1538-4357/aa9ff4}
  {\bibfield  {journal} {\bibinfo  {journal} {\apj}\ }\textbf {\bibinfo
  {volume} {852}},\ \bibinfo {eid} {97} (\bibinfo {year} {2018})},\ \Eprint
  {http://arxiv.org/abs/1707.09353} {arXiv:1707.09353} \BibitemShut {NoStop}%
\bibitem [{\citenamefont {{Rykoff}}\ \emph {et~al.}(2014)\citenamefont
  {{Rykoff}}, \citenamefont {{Rozo}}, \citenamefont {{Busha}}, \citenamefont
  {{Cunha}}, \citenamefont {{Finoguenov}}, \citenamefont {{Evrard}},
  \citenamefont {{Hao}}, \citenamefont {{Koester}}, \citenamefont
  {{Leauthaud}}, \citenamefont {{Nord}} \emph {et~al.}}]{rykoff14}%
  \BibitemOpen
  \bibfield  {author} {\bibinfo {author} {\bibfnamefont {E.~S.}\ \bibnamefont
  {{Rykoff}}}, \bibinfo {author} {\bibfnamefont {E.}~\bibnamefont {{Rozo}}},
  \bibinfo {author} {\bibfnamefont {M.~T.}\ \bibnamefont {{Busha}}}, \bibinfo
  {author} {\bibfnamefont {C.~E.}\ \bibnamefont {{Cunha}}}, \bibinfo {author}
  {\bibfnamefont {A.}~\bibnamefont {{Finoguenov}}}, \bibinfo {author}
  {\bibfnamefont {A.}~\bibnamefont {{Evrard}}}, \bibinfo {author}
  {\bibfnamefont {J.}~\bibnamefont {{Hao}}}, \bibinfo {author} {\bibfnamefont
  {B.~P.}\ \bibnamefont {{Koester}}}, \bibinfo {author} {\bibfnamefont
  {A.}~\bibnamefont {{Leauthaud}}}, \bibinfo {author} {\bibfnamefont
  {B.}~\bibnamefont {{Nord}}},  \emph {et~al.},\ }\href {\doibase
  10.1088/0004-637X/785/2/104} {\bibfield  {journal} {\bibinfo  {journal}
  {\apj}\ }\textbf {\bibinfo {volume} {785}},\ \bibinfo {eid} {104} (\bibinfo
  {year} {2014})},\ \Eprint {http://arxiv.org/abs/1303.3562} {arXiv:1303.3562}
  \BibitemShut {NoStop}%
\bibitem [{\citenamefont {{Rozo}}\ \emph {et~al.}(2016)\citenamefont {{Rozo}},
  \citenamefont {{Rykoff}}, \citenamefont {{Abate}}, \citenamefont {{Bonnett}},
  \citenamefont {{Crocce}}, \citenamefont {{Davis}}, \citenamefont {{Hoyle}},
  \citenamefont {{Leistedt}}, \citenamefont {{Peiris}}, \citenamefont
  {{Wechsler}} \emph {et~al.}}]{rozo15}%
  \BibitemOpen
  \bibfield  {author} {\bibinfo {author} {\bibfnamefont {E.}~\bibnamefont
  {{Rozo}}}, \bibinfo {author} {\bibfnamefont {E.~S.}\ \bibnamefont
  {{Rykoff}}}, \bibinfo {author} {\bibfnamefont {A.}~\bibnamefont {{Abate}}},
  \bibinfo {author} {\bibfnamefont {C.}~\bibnamefont {{Bonnett}}}, \bibinfo
  {author} {\bibfnamefont {M.}~\bibnamefont {{Crocce}}}, \bibinfo {author}
  {\bibfnamefont {C.}~\bibnamefont {{Davis}}}, \bibinfo {author} {\bibfnamefont
  {B.}~\bibnamefont {{Hoyle}}}, \bibinfo {author} {\bibfnamefont
  {B.}~\bibnamefont {{Leistedt}}}, \bibinfo {author} {\bibfnamefont {H.~V.}\
  \bibnamefont {{Peiris}}}, \bibinfo {author} {\bibfnamefont {R.~H.}\
  \bibnamefont {{Wechsler}}},  \emph {et~al.},\ }\href {\doibase
  10.1093/mnras/stw1281} {\bibfield  {journal} {\bibinfo  {journal} {\mnras}\
  }\textbf {\bibinfo {volume} {461}},\ \bibinfo {pages} {1431} (\bibinfo {year}
  {2016})},\ \Eprint {http://arxiv.org/abs/1507.05460} {arXiv:1507.05460
  [astro-ph.IM]} \BibitemShut {NoStop}%
\bibitem [{\citenamefont {Silk}(1968)}]{silk68}%
  \BibitemOpen
  \bibfield  {author} {\bibinfo {author} {\bibfnamefont {J.}~\bibnamefont
  {Silk}},\ }\href@noop {} {\bibfield  {journal} {\bibinfo  {journal} {\apj}\
  }\textbf {\bibinfo {volume} {151}},\ \bibinfo {pages} {459} (\bibinfo {year}
  {1968})}\BibitemShut {NoStop}%
\bibitem [{\citenamefont {{Seljak}}\ and\ \citenamefont
  {{Zaldarriaga}}(2000)}]{seljak00}%
  \BibitemOpen
  \bibfield  {author} {\bibinfo {author} {\bibfnamefont {U.}~\bibnamefont
  {{Seljak}}}\ and\ \bibinfo {author} {\bibfnamefont {M.}~\bibnamefont
  {{Zaldarriaga}}},\ }\href {\doibase 10.1086/309098} {\bibfield  {journal}
  {\bibinfo  {journal} {\apj}\ }\textbf {\bibinfo {volume} {538}},\ \bibinfo
  {pages} {57} (\bibinfo {year} {2000})},\ \Eprint
  {http://arxiv.org/abs/astro-ph/9907254} {astro-ph/9907254} \BibitemShut
  {NoStop}%
\bibitem [{\citenamefont {{Navarro}}\ \emph {et~al.}(1996)\citenamefont
  {{Navarro}}, \citenamefont {{Frenk}},\ and\ \citenamefont
  {{White}}}]{navarro96}%
  \BibitemOpen
  \bibfield  {author} {\bibinfo {author} {\bibfnamefont {J.~F.}\ \bibnamefont
  {{Navarro}}}, \bibinfo {author} {\bibfnamefont {C.~S.}\ \bibnamefont
  {{Frenk}}}, \ and\ \bibinfo {author} {\bibfnamefont {S.~D.~M.}\ \bibnamefont
  {{White}}},\ }\href {\doibase 10.1086/177173} {\bibfield  {journal} {\bibinfo
   {journal} {\apj}\ }\textbf {\bibinfo {volume} {462}},\ \bibinfo {pages}
  {563} (\bibinfo {year} {1996})},\ \Eprint
  {http://arxiv.org/abs/arXiv:astro-ph/9508025} {arXiv:astro-ph/9508025}
  \BibitemShut {NoStop}%
\bibitem [{\citenamefont {{Bartelmann}}(1996)}]{bartelmann96}%
  \BibitemOpen
  \bibfield  {author} {\bibinfo {author} {\bibfnamefont {M.}~\bibnamefont
  {{Bartelmann}}},\ }\href@noop {} {\bibfield  {journal} {\bibinfo  {journal}
  {\aap}\ }\textbf {\bibinfo {volume} {313}},\ \bibinfo {pages} {697} (\bibinfo
  {year} {1996})},\ \Eprint {http://arxiv.org/abs/arXiv:astro-ph/9602053}
  {arXiv:astro-ph/9602053} \BibitemShut {NoStop}%
\bibitem [{\citenamefont {{Oguri}}\ and\ \citenamefont
  {{Hamana}}(2011)}]{oguri11}%
  \BibitemOpen
  \bibfield  {author} {\bibinfo {author} {\bibfnamefont {M.}~\bibnamefont
  {{Oguri}}}\ and\ \bibinfo {author} {\bibfnamefont {T.}~\bibnamefont
  {{Hamana}}},\ }\href {\doibase 10.1111/j.1365-2966.2011.18481.x} {\bibfield
  {journal} {\bibinfo  {journal} {\mnras}\ }\textbf {\bibinfo {volume} {414}},\
  \bibinfo {pages} {1851} (\bibinfo {year} {2011})},\ \Eprint
  {http://arxiv.org/abs/1101.0650} {arXiv:1101.0650} \BibitemShut {NoStop}%
\bibitem [{\citenamefont {{Seljak}}(2000)}]{seljak00a}%
  \BibitemOpen
  \bibfield  {author} {\bibinfo {author} {\bibfnamefont {U.}~\bibnamefont
  {{Seljak}}},\ }\href {\doibase 10.1046/j.1365-8711.2000.03715.x} {\bibfield
  {journal} {\bibinfo  {journal} {\mnras}\ }\textbf {\bibinfo {volume} {318}},\
  \bibinfo {pages} {203} (\bibinfo {year} {2000})},\ \Eprint
  {http://arxiv.org/abs/arXiv:astro-ph/0001493} {arXiv:astro-ph/0001493}
  \BibitemShut {NoStop}%
\bibitem [{\citenamefont {{Cooray}}\ and\ \citenamefont
  {{Sheth}}(2002)}]{cooray02}%
  \BibitemOpen
  \bibfield  {author} {\bibinfo {author} {\bibfnamefont {A.}~\bibnamefont
  {{Cooray}}}\ and\ \bibinfo {author} {\bibfnamefont {R.}~\bibnamefont
  {{Sheth}}},\ }\href {\doibase 10.1016/S0370-1573(02)00276-4} {\bibfield
  {journal} {\bibinfo  {journal} {\physrep}\ }\textbf {\bibinfo {volume}
  {372}},\ \bibinfo {pages} {1} (\bibinfo {year} {2002})},\ \Eprint
  {http://arxiv.org/abs/astro-ph/0206508} {astro-ph/0206508} \BibitemShut
  {NoStop}%
\bibitem [{\citenamefont {{Oguri}}\ and\ \citenamefont
  {{Takada}}(2011)}]{oguri10}%
  \BibitemOpen
  \bibfield  {author} {\bibinfo {author} {\bibfnamefont {M.}~\bibnamefont
  {{Oguri}}}\ and\ \bibinfo {author} {\bibfnamefont {M.}~\bibnamefont
  {{Takada}}},\ }\href {\doibase 10.1103/PhysRevD.83.023008} {\bibfield
  {journal} {\bibinfo  {journal} {\prd}\ }\textbf {\bibinfo {volume} {83}},\
  \bibinfo {eid} {023008} (\bibinfo {year} {2011})},\ \Eprint
  {http://arxiv.org/abs/1010.0744} {arXiv:1010.0744} \BibitemShut {NoStop}%
\bibitem [{\citenamefont {{Rykoff}}\ \emph {et~al.}(2016)\citenamefont
  {{Rykoff}}, \citenamefont {{Rozo}}, \citenamefont {{Hollowood}},
  \citenamefont {{Bermeo-Hernandez}}, \citenamefont {{Jeltema}}, \citenamefont
  {{Mayers}}, \citenamefont {{Romer}}, \citenamefont {{Rooney}}, \citenamefont
  {{Saro}}, \citenamefont {{Vergara Cervantes}} \emph {et~al.}}]{rykoff16}%
  \BibitemOpen
  \bibfield  {author} {\bibinfo {author} {\bibfnamefont {E.~S.}\ \bibnamefont
  {{Rykoff}}}, \bibinfo {author} {\bibfnamefont {E.}~\bibnamefont {{Rozo}}},
  \bibinfo {author} {\bibfnamefont {D.}~\bibnamefont {{Hollowood}}}, \bibinfo
  {author} {\bibfnamefont {A.}~\bibnamefont {{Bermeo-Hernandez}}}, \bibinfo
  {author} {\bibfnamefont {T.}~\bibnamefont {{Jeltema}}}, \bibinfo {author}
  {\bibfnamefont {J.}~\bibnamefont {{Mayers}}}, \bibinfo {author}
  {\bibfnamefont {A.~K.}\ \bibnamefont {{Romer}}}, \bibinfo {author}
  {\bibfnamefont {P.}~\bibnamefont {{Rooney}}}, \bibinfo {author}
  {\bibfnamefont {A.}~\bibnamefont {{Saro}}}, \bibinfo {author} {\bibfnamefont
  {C.}~\bibnamefont {{Vergara Cervantes}}},  \emph {et~al.},\ }\href {\doibase
  10.3847/0067-0049/224/1/1} {\bibfield  {journal} {\bibinfo  {journal}
  {\apjs}\ }\textbf {\bibinfo {volume} {224}},\ \bibinfo {eid} {1} (\bibinfo
  {year} {2016})},\ \Eprint {http://arxiv.org/abs/1601.00621}
  {arXiv:1601.00621} \BibitemShut {NoStop}%
\bibitem [{\citenamefont {{McClintock}}\ \emph {et~al.}(2019)\citenamefont
  {{McClintock}}, \citenamefont {{Varga}}, \citenamefont {{Gruen}},
  \citenamefont {{Rozo}}, \citenamefont {{Rykoff}}, \citenamefont {{Shin}},
  \citenamefont {{Melchior}}, \citenamefont {{DeRose}}, \citenamefont
  {{Seitz}}, \citenamefont {{Dietrich}} \emph {et~al.}}]{mcclintock18}%
  \BibitemOpen
  \bibfield  {author} {\bibinfo {author} {\bibfnamefont {T.}~\bibnamefont
  {{McClintock}}}, \bibinfo {author} {\bibfnamefont {T.~N.}\ \bibnamefont
  {{Varga}}}, \bibinfo {author} {\bibfnamefont {D.}~\bibnamefont {{Gruen}}},
  \bibinfo {author} {\bibfnamefont {E.}~\bibnamefont {{Rozo}}}, \bibinfo
  {author} {\bibfnamefont {E.~S.}\ \bibnamefont {{Rykoff}}}, \bibinfo {author}
  {\bibfnamefont {T.}~\bibnamefont {{Shin}}}, \bibinfo {author} {\bibfnamefont
  {P.}~\bibnamefont {{Melchior}}}, \bibinfo {author} {\bibfnamefont
  {J.}~\bibnamefont {{DeRose}}}, \bibinfo {author} {\bibfnamefont
  {S.}~\bibnamefont {{Seitz}}}, \bibinfo {author} {\bibfnamefont {J.~P.}\
  \bibnamefont {{Dietrich}}},  \emph {et~al.},\ }\href {\doibase
  10.1093/mnras/sty2711} {\bibfield  {journal} {\bibinfo  {journal} {\mnras}\
  }\textbf {\bibinfo {volume} {482}},\ \bibinfo {pages} {1352} (\bibinfo {year}
  {2019})},\ \Eprint {http://arxiv.org/abs/1805.00039} {arXiv:1805.00039}
  \BibitemShut {NoStop}%
\bibitem [{\citenamefont {{Einasto}}\ and\ \citenamefont
  {{Haud}}(1989)}]{einasto89}%
  \BibitemOpen
  \bibfield  {author} {\bibinfo {author} {\bibfnamefont {J.}~\bibnamefont
  {{Einasto}}}\ and\ \bibinfo {author} {\bibfnamefont {U.}~\bibnamefont
  {{Haud}}},\ }\href@noop {} {\bibfield  {journal} {\bibinfo  {journal} {\aap}\
  }\textbf {\bibinfo {volume} {223}},\ \bibinfo {pages} {89} (\bibinfo {year}
  {1989})}\BibitemShut {NoStop}%
\bibitem [{\citenamefont {{Datta}}\ \emph {et~al.}(2018)\citenamefont
  {{Datta}}, \citenamefont {{Aiola}}, \citenamefont {{Choi}}, \citenamefont
  {{Devlin}}, \citenamefont {{Dunkley}}, \citenamefont {{D{\"u}nner}},
  \citenamefont {{Gallardo}}, \citenamefont {{Gralla}}, \citenamefont
  {{Halpern}}, \citenamefont {{Hasselfield}} \emph {et~al.}}]{datta18}%
  \BibitemOpen
  \bibfield  {author} {\bibinfo {author} {\bibfnamefont {R.}~\bibnamefont
  {{Datta}}}, \bibinfo {author} {\bibfnamefont {S.}~\bibnamefont {{Aiola}}},
  \bibinfo {author} {\bibfnamefont {S.~K.}\ \bibnamefont {{Choi}}}, \bibinfo
  {author} {\bibfnamefont {M.}~\bibnamefont {{Devlin}}}, \bibinfo {author}
  {\bibfnamefont {J.}~\bibnamefont {{Dunkley}}}, \bibinfo {author}
  {\bibfnamefont {R.}~\bibnamefont {{D{\"u}nner}}}, \bibinfo {author}
  {\bibfnamefont {P.~A.}\ \bibnamefont {{Gallardo}}}, \bibinfo {author}
  {\bibfnamefont {M.}~\bibnamefont {{Gralla}}}, \bibinfo {author}
  {\bibfnamefont {M.}~\bibnamefont {{Halpern}}}, \bibinfo {author}
  {\bibfnamefont {M.}~\bibnamefont {{Hasselfield}}},  \emph {et~al.},\ }\href
  {\doibase 10.1093/mnras/sty2934} {\bibfield  {journal} {\bibinfo  {journal}
  {\mnras}\ ,\ \bibinfo {pages} {2799}} (\bibinfo {year} {2018})},\ \Eprint
  {http://arxiv.org/abs/1811.01854} {arXiv:1811.01854 [astro-ph.CO]}
  \BibitemShut {NoStop}%
\bibitem [{\citenamefont {{Gupta}}\ \emph {et~al.}(2019)\citenamefont
  {{Gupta}}, \citenamefont {{Reichardt}}, \citenamefont {{Ade}}, \citenamefont
  {{Anderson}}, \citenamefont {{Archipley}}, \citenamefont {{Austermann}},
  \citenamefont {{Avva}}, \citenamefont {{Beall}}, \citenamefont {{Bender}},
  \citenamefont {{Benson}} \emph {et~al.}}]{gupta19}%
  \BibitemOpen
  \bibfield  {author} {\bibinfo {author} {\bibfnamefont {N.}~\bibnamefont
  {{Gupta}}}, \bibinfo {author} {\bibfnamefont {C.~L.}\ \bibnamefont
  {{Reichardt}}}, \bibinfo {author} {\bibfnamefont {P.~A.~R.}\ \bibnamefont
  {{Ade}}}, \bibinfo {author} {\bibfnamefont {A.~J.}\ \bibnamefont
  {{Anderson}}}, \bibinfo {author} {\bibfnamefont {M.}~\bibnamefont
  {{Archipley}}}, \bibinfo {author} {\bibfnamefont {J.~E.}\ \bibnamefont
  {{Austermann}}}, \bibinfo {author} {\bibfnamefont {J.~S.}\ \bibnamefont
  {{Avva}}}, \bibinfo {author} {\bibfnamefont {J.~A.}\ \bibnamefont {{Beall}}},
  \bibinfo {author} {\bibfnamefont {A.~N.}\ \bibnamefont {{Bender}}}, \bibinfo
  {author} {\bibfnamefont {B.~A.}\ \bibnamefont {{Benson}}},  \emph {et~al.},\
  }\href@noop {} {\bibfield  {journal} {\bibinfo  {journal} {arXiv e-prints}\
  ,\ \bibinfo {eid} {arXiv:1907.02156}} (\bibinfo {year} {2019})},\ \Eprint
  {http://arxiv.org/abs/1907.02156} {arXiv:1907.02156 [astro-ph.CO]}
  \BibitemShut {NoStop}%
\bibitem [{\citenamefont {{Carlstrom}}\ \emph {et~al.}(2002)\citenamefont
  {{Carlstrom}}, \citenamefont {{Holder}},\ and\ \citenamefont
  {{Reese}}}]{carlstrom02}%
  \BibitemOpen
  \bibfield  {author} {\bibinfo {author} {\bibfnamefont {J.~E.}\ \bibnamefont
  {{Carlstrom}}}, \bibinfo {author} {\bibfnamefont {G.~P.}\ \bibnamefont
  {{Holder}}}, \ and\ \bibinfo {author} {\bibfnamefont {E.~D.}\ \bibnamefont
  {{Reese}}},\ }\href {\doibase 10.1146/annurev.astro.40.060401.093803}
  {\bibfield  {journal} {\bibinfo  {journal} {Annual Review of Astronomy and
  Astrophysics}\ }\textbf {\bibinfo {volume} {40}},\ \bibinfo {pages} {643}
  (\bibinfo {year} {2002})},\ \Eprint {http://arxiv.org/abs/astro-ph/0208192}
  {arXiv:astro-ph/0208192 [astro-ph]} \BibitemShut {NoStop}%
\bibitem [{\citenamefont {{Hall}}\ and\ \citenamefont
  {{Challinor}}(2014)}]{hall14}%
  \BibitemOpen
  \bibfield  {author} {\bibinfo {author} {\bibfnamefont {A.}~\bibnamefont
  {{Hall}}}\ and\ \bibinfo {author} {\bibfnamefont {A.}~\bibnamefont
  {{Challinor}}},\ }\href {\doibase 10.1103/PhysRevD.90.063518} {\bibfield
  {journal} {\bibinfo  {journal} {\prd}\ }\textbf {\bibinfo {volume} {90}},\
  \bibinfo {eid} {063518} (\bibinfo {year} {2014})},\ \Eprint
  {http://arxiv.org/abs/1407.5135} {arXiv:1407.5135 [astro-ph.CO]} \BibitemShut
  {NoStop}%
\bibitem [{\citenamefont {{Yasini}}\ and\ \citenamefont
  {{Pierpaoli}}(2016)}]{yasini16}%
  \BibitemOpen
  \bibfield  {author} {\bibinfo {author} {\bibfnamefont {S.}~\bibnamefont
  {{Yasini}}}\ and\ \bibinfo {author} {\bibfnamefont {E.}~\bibnamefont
  {{Pierpaoli}}},\ }\href {\doibase 10.1103/PhysRevD.94.023513} {\bibfield
  {journal} {\bibinfo  {journal} {\prd}\ }\textbf {\bibinfo {volume} {94}},\
  \bibinfo {eid} {023513} (\bibinfo {year} {2016})},\ \Eprint
  {http://arxiv.org/abs/1605.02111} {arXiv:1605.02111 [astro-ph.CO]}
  \BibitemShut {NoStop}%
\bibitem [{\citenamefont {{Bender}}\ \emph {et~al.}(2018)\citenamefont
  {{Bender}}, \citenamefont {{Ade}}, \citenamefont {{Ahmed}}, \citenamefont
  {{Anderson}}, \citenamefont {{Avva}}, \citenamefont {{Aylor}}, \citenamefont
  {{Barry}}, \citenamefont {{Basu Thakur}}, \citenamefont {{Benson}},
  \citenamefont {{Bleem}} \emph {et~al.}}]{bender18}%
  \BibitemOpen
  \bibfield  {author} {\bibinfo {author} {\bibfnamefont {A.~N.}\ \bibnamefont
  {{Bender}}}, \bibinfo {author} {\bibfnamefont {P.~A.~R.}\ \bibnamefont
  {{Ade}}}, \bibinfo {author} {\bibfnamefont {Z.}~\bibnamefont {{Ahmed}}},
  \bibinfo {author} {\bibfnamefont {A.~J.}\ \bibnamefont {{Anderson}}},
  \bibinfo {author} {\bibfnamefont {J.~S.}\ \bibnamefont {{Avva}}}, \bibinfo
  {author} {\bibfnamefont {K.}~\bibnamefont {{Aylor}}}, \bibinfo {author}
  {\bibfnamefont {P.~S.}\ \bibnamefont {{Barry}}}, \bibinfo {author}
  {\bibfnamefont {R.}~\bibnamefont {{Basu Thakur}}}, \bibinfo {author}
  {\bibfnamefont {B.~A.}\ \bibnamefont {{Benson}}}, \bibinfo {author}
  {\bibfnamefont {L.~S.}\ \bibnamefont {{Bleem}}},  \emph {et~al.},\ }in\ \href
  {\doibase 10.1117/12.2312426} {\emph {\bibinfo {booktitle} {Millimeter,
  Submillimeter, and Far-Infrared Detectors and Instrumentation for Astronomy
  IX}}},\ \bibinfo {series} {Society of Photo-Optical Instrumentation Engineers
  (SPIE) Conference Series}, Vol.\ \bibinfo {volume} {10708}\ (\bibinfo {year}
  {2018})\ p.\ \bibinfo {pages} {1070803},\ \Eprint
  {http://arxiv.org/abs/1809.00036} {arXiv:1809.00036 [astro-ph.IM]}
  \BibitemShut {NoStop}%
\end{thebibliography}%
\end{document}